\documentclass{aa}  
\usepackage[T1]{fontenc}
\usepackage{graphicx}
\usepackage{hyperref}
\usepackage{amsmath}    % Advanced maths commands
\usepackage{amssymb}    % Extra maths symbols
\usepackage{multirow} 
\usepackage{float}
\usepackage{url}
\usepackage{booktabs} 
\usepackage{rotating}
\usepackage{adjustbox}
\usepackage{comment}
\usepackage{ragged2e}
\usepackage{newtxtext,newtxmath}  % Prefer this over txfonts
\usepackage{color}
\usepackage{verbatim}
\usepackage[switch]{lineno}
\usepackage[section]{placeins}
\usepackage{longtable}

\begin{document}

%\linenumbers

   \title{New insight into the variability of the Be star $\pi$~Aquarii: Determination of stellar and disk parameters}

   \author{D. Concha\inst{1,2}, C. Arcos\inst{2}, D. Turis-Gallo\inst{2}, T. B. Souza\inst{2}, M. Cur\'e\inst{2}, R. S. Levenhagen\inst{3} \and I. Araya\inst{4}}
   %\fnmsep\thanks{Just to show the usage of the elements in the author field}

    \institute{Instituto de Astronom\'ia y Ciencias Planetarias de Atacama, Universidad de Atacama, Copayapu 485, Copiap\'o, Chile. \\
    \email{david.concha@postgrados.uda.cl}
    \and
            Instituto de F\'isica y Astronom\'ia, Facultad de Ciencias, Universidad de Valpara\'iso,  Av. Gran Bretana 1111, Valpara\'iso, Chile.
    \and
            Departamento de Física, Universidade Federal de Sao Paulo, Rua Prof. Artur Riedel, 275, 09972-270, Diadema, SP, Brazil.
    \and
            Centro Multidisciplinario de F\'isica, Vicerrector\'ia de Investigaci\'on, Universidad Mayor, 8580745 Santiago, Chile
            % \thanks{The university of heaven temporarily does not accept e-mails}
             }

   \date{Received XXX, 2024; accepted month day, yyyy}

% \abstract{}{}{}{}{} 
% 5 {} token are mandatory
%% AMERICAN ENGLISH 
 
  \abstract
{The Be star $\pi$~Aquarii shows a peculiar X-ray emission ($\gamma$ Cas type) commonly attributed to its outer disk interacting with a low-mass companion star, probably a white dwarf.}
   {In this work, we study the long-term variability of the optical spectra of $\pi$~Aquarii in order to derive the stellar and disk parameters of specific epochs when the disk shows major changes.}
   {We identified several spectral lines and studied specific observation dates for Balmer, helium, silicon, and iron emission lines in detail. We determined the stellar parameters using stellar atmosphere models, considering the oblate geometry due to the fast rotation, and we derived the disk parameters through H$\alpha$ modeling and considering a viscous decretion disk parametric model.}
   {We found the following stellar parameters: $\rm M=(11.0\pm1.9)\ M_\odot$, $\rm <\log g> = (4.03\pm0.10)\ dex$, $\rm Rp = (5\pm0.8)\ R_\odot$, $\rm R_{eq} = (1.15\pm0.088)\ R_p$, $\rm T_p = (25000\pm1174)\ K$, $\rm v\sin i = (271\pm13)\ km/s$, and a $\omega = 0.82\pm0.05$. Over five epochs, the H$\alpha$ emission line evolved from shell profiles to double-peaked to triple-peaked to flat-topped structures. We found that the density in the disk decays faster (with n $\sim$ 3.9) on December 22, 2001. In this epoch, the H$\alpha$ emission line presented a low-intensity shell profile modeled with a low base surface density ($\Sigma_0 = 0.01 \rm g cm^{-2}$), indicating a "small" disk. Then, from 2011 to (September) 2014, the disk decreased with a slowly decaying density distribution (n $\sim$ 3.2). The disk grew drastically from October 2014 until November 2022, increasing 18 times its initial H$\alpha$ equivalent width value. At that time, we estimated an emitting region of $\sim 65 R_{\odot}$. We also measured the inclination angle variation in $\sim$ 10$^{\circ}$ in 20 years, with the results indicating a likely precessing disk. We found that FeII 5018 $\AA$ \, covers an emission region larger than H$\alpha$ and is the only FeII emission line showing a different shape profile than the rest of the lines.}
   {We assert that the disk of $\pi$ Aquarii is in a misaligned binary system going through shell profiles to double-peak separation to triple-peaked to flat-topped profiles, and it now shows an asymmetric double-peak separation profile again. The emission line FeII 5018 $\AA$ indicates changes in the outer disk probably related to the white dwarf. We propose that the white dwarf cross the Be disk at two points in its orbit, and it is at these moments that it captures material and temporarily increases its X-ray emission.}


\titlerunning{Variability of the Be star $\pi$~Aquarii}\authorrunning{Concha,D et al.}
\maketitle
\section{Introduction}
 The star $\rm \pi$~Aquarii (HD~212571, HIP~110672) is a bright rapidly rotating classical Be star (B1III-IVe) presenting mid- and long-term variability. Notably, the star has been observed for more than 100 years. \cite{McLaughlin1962} reported variations over 50 years in the line profile of H$\alpha$ and H$\gamma$, where more of the profiles were observed in double-peaked line emission, probably indicating the active phase of the disk of $\pi$~Aquarii. He also noted a significant change from 0.5 to 4 units in the violet-to-red peak intensity ratio (V/R), pointing out a signal of asymmetries in the disk. This was the first attempt to study the disk activity in $\pi$~Aquarii, even though the stellar and disk parameters were undetermined.

Several authors \citep{Nordh1974,Underhill1979,Snow1981,Goraya1985,Theodossiou1985,Fabregat1990,Zorec1991} have studied this star over the years, trying to establish the fundamental parameters by using different observational techniques, such as multi-color photometry, interferometry, polarimetry, and high-resolution spectroscopy. In summary, based on all of these works, the effective temperature, T$_{\rm eff}$, ranges between 22500 and 30000 K; the effective gravity, $\log g$, is between 3.3 and 4.0 dex; the terminal velocity of the wind, v$_{\infty}$, has been calculated as $\sim$ 1450 km\,s$^{-1}$; and the mass-loss rate is $\dot{\rm M}\sim$~2.61~$\times$~10$^{-9}$\ M$_\odot$ yr$^{-1}$ \citep{Snow1981}.

\cite{McDavid1986} used polarimetric data and found that the previous active phase started approximately in 1950, reaching its maximum around 1985, where the polarization level was higher than 1$\%$. After that, as the emission line decreased, so did the brightness of the central star, reaching a minimum value in 1996 and extending its quasi-normal phase until 2002. \cite{Hanuschik1996} proposed that $\pi$~Aquarii has a companion due to the broad and complex H$\alpha$ line profile observed. 
\cite{Bjorkman2002} performed long-term observations of the star, obtaining information about the behavior of the disk phases of $\pi$~Aquarii. They calculated new fundamental parameters of the binary system for the first time by fitting the photometric data with Kurucz local temperature equilibrium (LTE) stellar atmosphere models \citep{Kurucz1994}. The best fit they obtained had the following values: $\rm M_1 = (11 \pm 1.5)\,M_{\odot}$, $\rm T_{eff} = (24000 \pm 1000)\,K$, and $\rm A_V = (0.15 \pm 0.03)\, mag$. They used a distance of $340_{-70}^{+105}\,\rm pc$ \citep{Hipparcos1997}, $\rm \log\left(\rm L_{bol}/L_{\odot}\right) = (4.1 \pm 0.3)$ dex, and a $\rm R_{\star} = (6.1 \pm 2.5)\, R_{\odot}$. They also suggested an inclination angle in the range $50^{\circ} \leq i \leq 75^{\circ}$ based on previous studies on the FeII~5317 \AA \, double-peaked emission line and taking into account that the central depression observed in Balmer lines between 1989 and 1995 is not as deep as it appears if seen edge-on. By examining the periodical radial velocity (RV) variation of the H$\alpha$ emission line, they also concluded that if $\pi$~Aquarii is a spectroscopic binary, a gas envelope should surround the companion. Considering this fact, they derived the following binary parameters: $\rm
M_1\sin^3{i} = 12.4\ M_{\odot}$, $\rm M_2\sin i ^3 = 2.0\ M_{\odot}$, $\rm a=0.96\ \arcsin{i}\ AU$, and $\rm M_2 = 0.16\ M_1$.

\cite{Zharikov2013} recollected data for 40 orbital cycles from 2004 to 2013 and analyzed variations of double-peak asymmetries in H$\alpha$ profiles. They searched for a periodicity in the V/R, finding 84.1 days, which agrees with the orbital period of the binary companion of $\pi$~Aquarii \citep{Bjorkman2002}. In addition, they used evolutionary models to calculate the stellar parameters of the binary system. They obtained $\rm M_{1} = (14\pm1)\ M_\odot$, $\rm M_2 = (2.31\pm16)\ M_\odot$, $\rm T_{eff} = (24000\pm1000)\ K$, $\rm \log L/L_\odot = (4.7\pm1)\ dex$, $\rm R_1 = (13\pm1.4)\ R_\odot$, mass ratio $\rm q \geq 0.165$, $\rm 65^{\circ} \le i \le 85^{\circ}$, and $\rm a=0.96$ AU, with a distance of $\rm 740\pm90$ pc, which is more than twice the previously reported value. They studied the disk's structure by employing Doppler tomography, which showed a non-uniform structure. The structure they observed consists of a ring with inner and outer radii (in velocity units) of 450 and 200 km s$^{-1}$ and an extended stable region. The brightness of this extended region presented an S-wave that appeared similar to a one-armed spiral density pattern structure in phase with the secondary. In addition, the computation of the disk radius of $\rm \sim 65\,R_{\odot}$ ($\rm \sim 0.33$ AU) was estimated, assuming the Keplerian motion of a particle in a circular orbit at the outer disk.

\cite{Naze2019a} collected data between 2013 and 2019 from the Be Star Spectra (BeSS) database\footnote{\url{http://basebe.obspm.fr/basebe/}} \citep{Neiner2011} and found that the emission line profile seen in H$\alpha$ almost disappears in January 2014. Then, the emission resembles the active phase seen from 1950 to 1990. A striking result was that the extended stable region (observed by \citeauthor{Zharikov2013}) is no longer present after 2016, proving a change in the density distribution of the disk. In addition, the disk's inner edge was closer (in velocity) to the central star, suggesting an increased disk size. 

Observations from space-based photometric data have shown that pulsation is a common feature of classical Be stars, with these stars primarily pulsating in low-order gravity g modes, where gravity serves as the restoring force. Internal gravity waves transport angular momentum from the stellar interior to the envelope, causing an increment in the stellar rotation. This could create the conditions necessary to initiate an outburst, during which mass and angular momentum are transferred to the disk \citep{Labadie-Bartz2022}. \cite{Naze2020} analyzes the photometric data of $\pi$~Aquarii, revealing low-amplitude, coherent variability at high frequencies, suggesting p modes (pressure acts as the restoring force). This is atypical for hot Be stars because their internal structure and high rotation rates create conditions that do not favor the formation of stable pressure gradients in the outer layers of the star. However, other classical Be stars are also observed to have p mode pulsations \citep{Walker2005,Huat2009,Labadie-Bartz2020}.

%\cite{Labadie-bartz2022} analyzed the light curve of 432 classical Be stars from The NASA Transiting Exoplanet Survey Satellite (TESS), finding that 87\% of the samples show one or more groups of closely spaced frequencies, and almost all samples have multiple periodic signals between 0.5 and 4 day$^{-1}$. As a reference, frequencies lower than 6 day${^-1}$ are typical for g-modes, while between 6 and 15 day$^{-1}$ are common for p-modes (pressure being the restoring force). 

Using observations with the spatial telescope XMM-Newton, \cite{Naze2017} studied the variable thermal X-ray emission of $\pi$~Aquarii, classifying it as a $\gamma$ Cas type. These are a subgroup of classical Be stars with hard (more than 5–10 keV) and moderately strong continuous thermal X-ray flux.  It has been suggested that the X-ray emission originates from accretion onto a secondary star.
To determine the source of the unusual X-ray emission, simultaneous optical and X-ray observations were performed by \cite{Naze2019b} from April to December 2018. In their comparison, the equivalent width (EW) of the H$\alpha$ line increased from -22 to -24.5\,\AA\,, in contrast to the X-ray fluxes, which do not display correlation with time nor the EW of the H$\alpha$ line. They note that although H$\alpha$ has undergone dramatic changes, no significant changes were observed in the X-ray emission level, which is contrary to what has been found in other $\gamma$ Cas stars. Moreover, they concluded that the behavior of $\pi$~Aquarii points to a missing ingredient in the scenarios they considered. \\
%(this year, the star presented a strong H$\alpha$ emission). In contrast to what they previously found,  the RV amplitude is half of that measured by \cite{Bjorkman2002}. 
Regarding the companion of $\pi$~Aquarii, \cite{Tsujimoto2023} proposed that the secondary star is a non-magnetic or magnetic accreting white dwarf (WD). This information was obtained through two methods. (i) By comparing effective temperature and stellar luminosity to evolutionary models, they obtained an evolutionary mass of 10 M$_\odot$ (for the Be star). (ii) By giving a range of possible mass for the Be star (9 to 15 M$_\odot$) and for the inclination angle (50 to 75$^{\circ}$), they obtained a dynamical mass for the secondary between $\sim$ 0.8 and 1.4 M$_\odot$. Other works support the origin of X-rays in $\gamma$ Cas stars due to accretion onto WD companions \citep{2023Gies,2024Klement}.

In this work, we study the line profile variability of several spectral lines identified in the spectra of $\pi$~Aquarii in order to characterize the behavior of the star and the disk. We measure the EW, V/R, and double-peak separation (DPS) of the emission lines. For this study, we used spectroscopy data from the Be Stars Observation Survey (BeSOS) and BeSS databases (described in the next section). We derived a period from the V/R. We also obtained the stellar parameters under consideration of the oblateness of the star (limb and gravity darkening) and characterized the changes in the disk (density, inclination, etc.) by using the BeATLAS grid and Bayesian methods.

\section{Observational data}
The BeSOS database\footnote{\url{https://besos.ifa.uv.cl/}}\citep{Arcos2018} contains optical spectra of southern Be stars, observed with an
echelle spectrograph of R $\sim$ 17000. The catalog is complete up to magnitude V$\sim$ 11. For each star, the catalog gathers information about its parameters,
such as coordinates, magnitude, temperature, gravity, projected rotation velocity, and spectral type, among others. A total of 7 spectra of $\pi$~Aquarii are
available from the BeSOS database, where two were observed on November 14, 2012; three on July 24, 2013; and the last two on October 24, 2015. The spectra were averaged
each night, obtaining three spectra with an effective wavelength range of 4200 to 7200 \AA. 

The Be Star Spectra database \citep{Neiner2011} provides a comprehensive catalog of classical Be stars, Herbig Ae/Be stars, and B[e] supergiants. This database compiles spectra collected by both professional and amateur astronomers. Specifically, for $\pi$ Aquarii, we have a total of 478 spectra from 2001 to 2024, covering only the H$\alpha$ wavelength region and a total of 16 spectra covering the optical region. These include a spectrum from 2001 and at least one spectrum per year from 2008 to 2024. The wavelength range varies for each spectrum, as detailed in Table~\ref{tab:bess_combined}. 

\begin{figure}[htbp]
  \resizebox{\hsize}{!}{\includegraphics{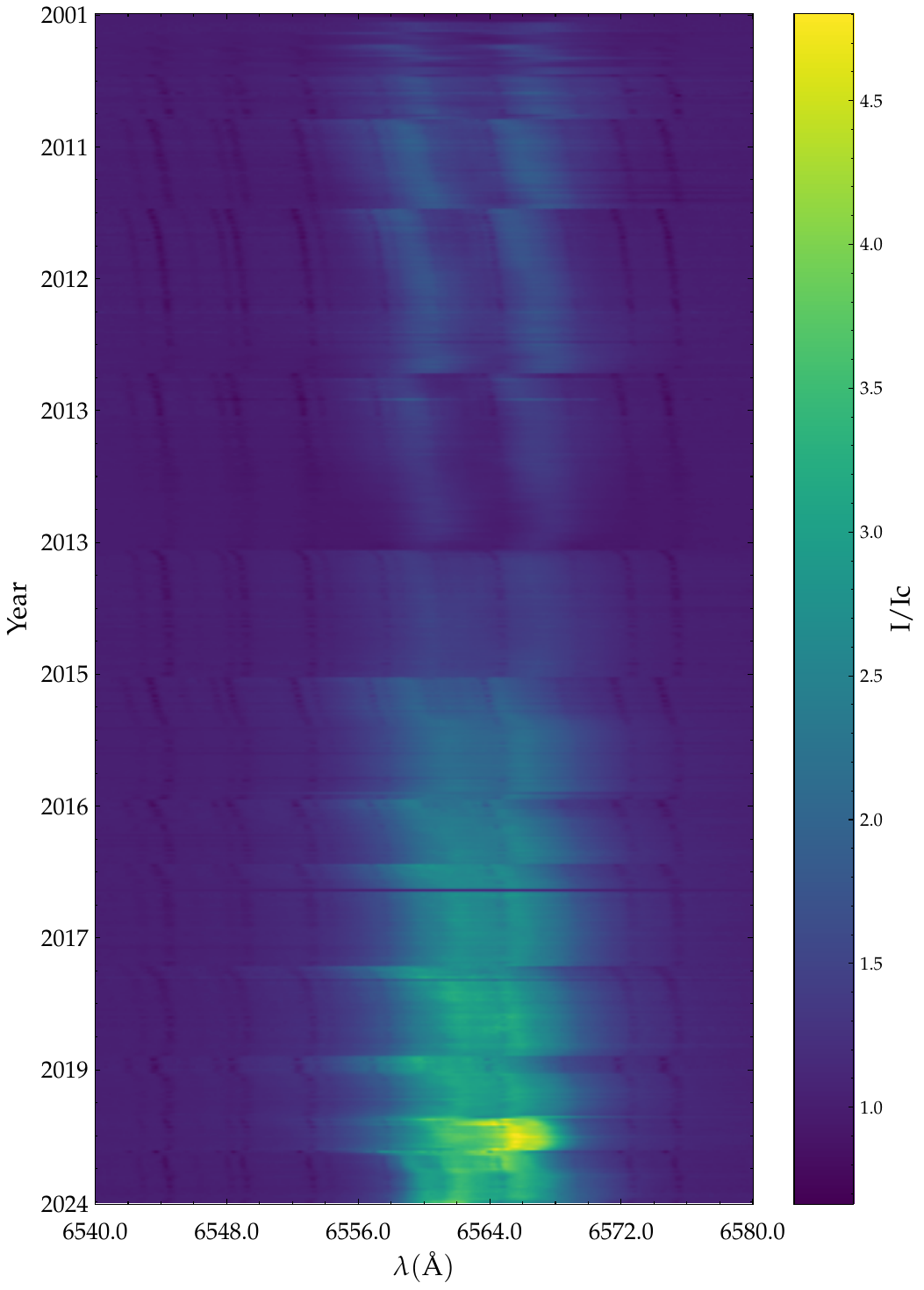}}
  \resizebox{\hsize}{!}{\includegraphics{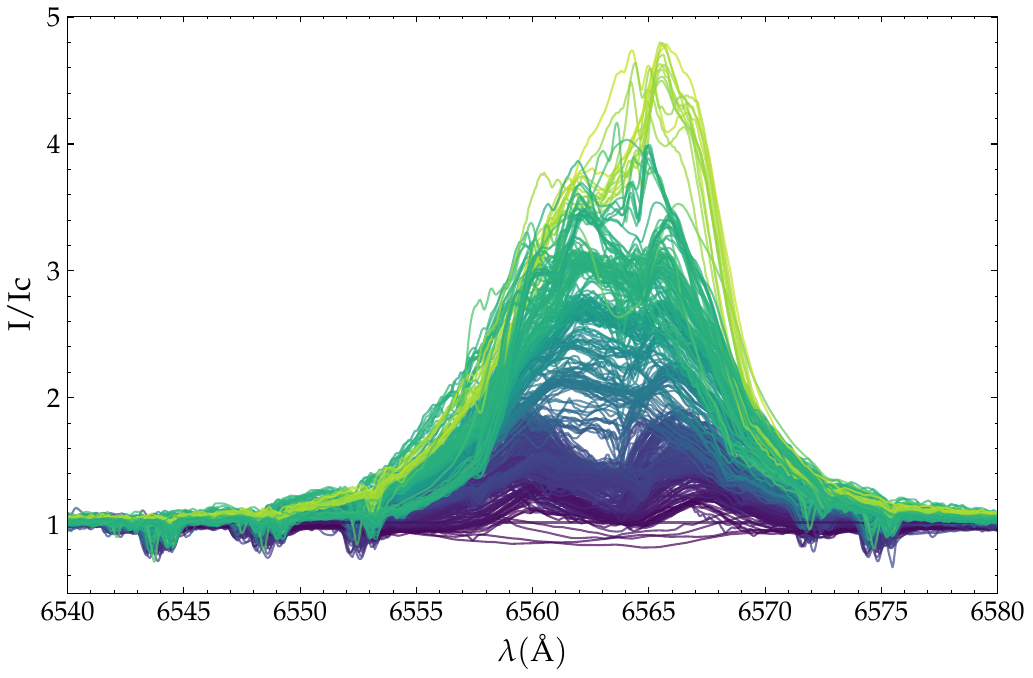}}
  \caption{Dynamical plot of H$\alpha$ emission line profiles observed between 2001 and 2024.}
  \label{fig:multiobs}
\end{figure}

Figure~\ref{fig:multiobs} contains all H$\alpha$ emission line profiles plotted to appreciate the variation over the years, where epochs are sorted from the top to the bottom. This figure shows the high variation of this star over the years. Lighter colors (yellow) indicate higher intensity values. A great difference is observed between 2015 and 2024, where the intensity of the line increases considerably compared to the previous years. Details about this observed variation are presented in the results section under the EW, V/R, and DPS variability.

\section{Method}
Several spectral lines were identified in the available spectra for $\pi$ Aquarii. Some lines constantly appear in absorption or emission, while others exhibit both behaviors. Table~\ref{tab:lines} provides a detailed overview of these identified spectral lines, including descriptions of weak and strong lines, as well as variations such as double-peaked profiles.
The following sections outline the methodology used to determine the stellar parameters and measurements on several lines and derive the disk parameters.

\begin{table*}
\caption{Observed spectral lines in $\pi$ Aquarii and their variations.}
\label{tab:lines}
\centering
\small
\begin{tabular}{cc cc cc cc}
\hline\hline
\noalign{\smallskip}
\multicolumn{2}{c}{He I} & \multicolumn{2}{c}{Fe II} & \multicolumn{2}{c}{Balmer Lines} & \multicolumn{2}{c}{Other Lines}  \\
\hline
\noalign{\smallskip}
Line & Description & Line & Description & Line & Description & Line & Description  \\
\hline
\noalign{\smallskip}
4388 & $\blacklozenge$ & 4233 & $\bigcirc$ & H$\delta$ 4102 & $\blacktriangle$  & O II 4254\tablefootmark{?} & $\blacklozenge$ \\
4471 & $\blacklozenge$ & 4549\tablefootmark{a} & $\bigotimes$ & H$\gamma$ 4340 & $\blacktriangle$  & O II 4417 & $\blacklozenge$ \\
4713 & $\blacklozenge$ & 4584 & $\bigcirc$ & H$\beta$ 4861 & $\blacktriangle$ & Mg II 4481 &$\blacklozenge$ \\
4921 & $\blacklozenge$ & 4629 & $\bigotimes$ & H$\alpha$ 6563 & $\bullet$,$\blacktriangle$  & Si III 4552\tablefootmark{a} & $\bigotimes$  \\
5015\tablefootmark{a} & $\bullet$ & 4667 & $\bigotimes$ &  &  & Si III 4567 & $\bigotimes$ \\
5047 & $\blacklozenge$ & 5018\tablefootmark{a} & $\bullet$ &  &  & C III/O II 4650\tablefootmark{?} & $\blacklozenge$  \\
5875 & $\bullet$ & 5169 & $\bigcirc$ &  &  & Ti I/O II 4675\tablefootmark{?} & $\blacklozenge$\\
6678 & $\circ$ & 5198 & $\bigcirc$ &  &  & Si II 6347\tablefootmark{a} & $\bigotimes$ \\
7065 & $\circ$ & 5235 & $\bigcirc$ &  &  & Si II 6371\tablefootmark{a} & $\bigotimes$ \\
     &  & 5276 & $\bigcirc$ &  &  &   & \\
     &  & 5316 & $\bigcirc$ &  &  &  & \\
     &  & 5363 & $\bigcirc$ &  &  &  & \\
     &  & 6249 & $\bigotimes$ &  &  &  & \\
     &  & 6369\tablefootmark{a} & $\bigotimes$ &  &  &  & \\
     &  & 6385 & $\bigotimes$ &  &  &  & \\
\hline
\end{tabular}
\tablefoot{
\tablefoottext{?}{Uncertainty about the element.}
\tablefoottext{a}{Overlapping lines.}
}
\tablefoot{
Description of each line: 
$\bigotimes$ No signal or very weak transition between emission and absorption lines. 
$\bullet$ DPS emission with variations over the years. 
$\blacklozenge$ Absorption line with some variations. 
$\circ$ DPS in early years (2001-2014), disappearing and reforming in later years. 
$\blacktriangle$ Transition between shell-type and DPS profiles. 
$\bigcirc$ DPS formation in final years (2014-2024).
}
\end{table*}

\subsection{Determination of stellar parameters}\label{sec:meth:sp}
The spectra of Be stars in the active phase are contaminated by emissions from the ionized disk, affecting not only hydrogen lines but also helium and other metallic lines. Therefore, to determine the stellar parameters, we chose the deepest absorption line, HeI~4471~\AA \ (July 23, 2013), among the optical data available covering that range. The HeI~4471~\AA \, is one of the most intense line profiles within all helium lines in the optical spectrum of O and B stars and is mainly constituted by a blend of a triplet $2^3P - 4^3D$ and a forbidden $2^3P - 4^3F$ transitions. As the radiation field of OB stars is not intense enough to populate the upper level of this transition \citep{Osterbrock1989}, this line is, in principle, less perturbed by the contribution of the circumstellar environment. We used a new Python version of the \texttt{ZPEKTR} code \citep{Levenhagen2011,Levenhagen2014,2024Leven}, which adopts a grid of non-LTE spectra as input, built up with the \texttt{SYNSPEC} code \citep{Hubeny1995}, and based on the stellar atmospheres models calculated with the \texttt{TLUSTY} code \citep{Hubeny1988}. The \texttt{ZPEKTR} code receives as input the radius $\rm R/R_{\odot}$, the mass $\rm M/M_{\odot}$ and the effective temperature $\rm T_{eff}$ of the unperturbed, non-rotating spherical star and also the stellar rotation rate $\rm v/v_c$ and its inclination angle $i$. The star is divided into hundreds of thousands of grid elements, which are rotated at the specified input velocity rate and geometrically deformed and tilted. Each mesh element on the stellar surface has its particular radius value, computed with a Newton-Raphson iteration scheme, and its particular set of local physical parameters $(\rm T_{eff}, \log{g}, v)$, which are evaluated obeying the classical von Zeipel's relation with index $\beta = 0.25$. In the next step, the code performs the numerical interpolation, within the mesh, of all local specific intensities using the non-LTE grid of \texttt{TLUSTY/SYNSPEC} models, accounting for a desired limb-darkening relation, and Doppler shifts these intensities accordingly. The output averaged flux spectrum is obtained by the numerical integration of all specific intensities emerging from the visible surface elements and brings the signatures of the effects of gravity darkening, limb darkening, and oblateness of the star. For this work, we used the linear Limb-darkening law \citep{Klinglesmith1970} and the Espinosa-Lara prescription for gravity darkening \citep{EspinosaLara2011,EspinosaLara2014}. We created a grid of 11294 models with \texttt{ZPEKTR} code to seek the best representation of the stellar parameters. The range of parameters explored is described in Table~\ref{tab:grid}, where the values are distinguished between the pole (subscript "p") and the equator (subscript "eq").

\begin{table}
\caption{Input parameters to create a grid of models using the \texttt{ZPEKTR} code.}
\label{tab:grid}
\centering
\begin{tabular}{l c c c}
Parameters & Initial & Final & Step \\ 
\hline\hline
\noalign{\smallskip}
$\rm M \, (M_{\odot})$ & 9 & 15 & 1 \\
$\rm R_{p} \, (R_{\odot})$ & 5 & 10 & 1 \\
$\rm T_{p} \, (\mathrm{kK})$ & 22 & 28 & 1 \\
$\rm i \, (^{\circ})$ & 50 & 75 & 5 \\
$\rm R_{eq} \, (R_{p})$ & 1.15 & 1.45 & 0.045 \\ 
\hline
N models & \multicolumn{3}{c}{11294} \\
\hline
\end{tabular}
\end{table}

To select the best models, we conducted a Chi-square statistical test comparing the synthetic line generated using the \texttt{ZPEKTR} code with the observed spectral line. This analysis focused on the spectral range encompassing the HeI 4471 Å line. The determination of the "best fit" models was based on examining the change in the slope of the Chi-square values plotted against the number of models considered.

\subsection{Estimation of equivalent widths}
Before measuring the EW, all spectra were normalized to the continuum. We used different methods depending on whether the observed line is in emission or absorption. In the case of emission lines, that is, Balmer and some HeI lines ($\lambda$ 5015, 5047, 5875, and 6678 \AA) and some FeII lines ($\lambda$  4232, 4584, 5169, 5198, 5235, 5276, 5316, and 5363 \AA), we subtracted the contribution of the stellar photosphere by using the result obtained with \texttt{ZPEKTR}. % (see Fig.~\ref{fig:ewem} as an example). 
After removing the stellar photosphere contribution, the EW was computed considering the zero moments (M$_0$). This method quantifies the width
(x$_i$) of a hypothetical rectangular line (of flux F$_i$) with a unit amplitude that has the same integrated area as the observed spectral
line\footnote{$M_0 = - \sum F_i\Delta x_i$} \citep{Balona1996}. We note that while subtracting the stellar contribution to the emission line, the
continuum is set to zero, and then we do not subtract the unit from the resulting value. Despite this, we added a negative sign to follow the formal
definition of EW for emission lines; that is, the more negative the EW, the stronger the line emission.
To estimate the uncertainties of the EWs, the Bootstrap re-sampling method \citep{Efron1979} was implemented. The method constructs hypothetical data sets
derived from the observed data, picking random points with replacement. Any data points may occur one or more times or may not be selected. The data points
are size N, which is equivalent to the size of the observed data. %The EWs are computed using Eq.~\ref{eq:EWemision} and \ref{eq:EWabs} with the new bootstrapped data set. 
The process was repeated 1000 times. The final value is given in the form: EW $\pm\,\sigma^{*}$, where $\sigma^{*}$ is the standard deviation of the entire bootstrapped data set.

\subsection{Violet to red peak intensity ratio and double-peak separation} 
\begin{figure*}
\begin{center}
\includegraphics[scale=0.5]{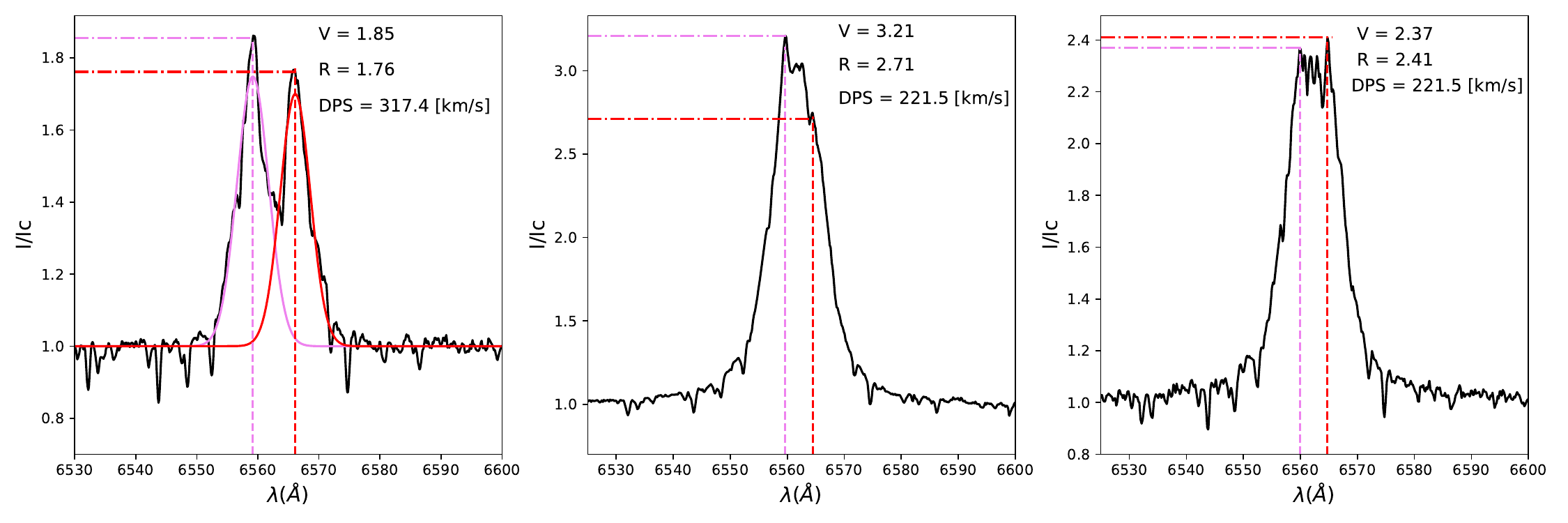}
\caption{Method used to obtain V/R and DPS for different shape profiles (solid black line). Case 1 (left panel): Distinguished DPS profile. The Gaussian fits are represented in pink and red for the violet and red peaks, respectively.  The dashed lines indicate the Gaussian fit's center and the values for both peak intensities. Values for this example are in the legend of the plot. Case 2 (middle panel): The code selects the maximum intensities from each side of the center (dashed vertical lines). Case 3 (right panel): Flat emission line. For these cases, the peaks were selected by visual inspection from each side of the geometrical center of the line (dashed vertical lines)}
\label{fig:cases}
\end{center}
\end{figure*}

For all emission lines shown in Table~\ref{tab:lines}, we calculated the V/R, which is the ratio between the violet and red emission peaks of circumstellar emission lines \citep{2009Stefl,2009Carciofi}. The DPS is the distance between the wavelength points of the violet and red emission peaks. The H$\alpha$ line profiles present a complex geometry over time. Some profiles present a flat shape in the peak intensity, while others present a third peak in emission. To achieve this calculation, we classified the observed emission lines in three different shapes (cases) described as follows: 

\begin{itemize}
    \item Case 1: This case is for well-distinguished DPS profiles. An example of this type of H$\alpha$ profile is shown in the left panel of Fig.~\ref{fig:cases}. The method was to fit two Gaussian profiles, one for each peak. The center of each Gaussian gives the violet and red peak values, respectively.
    \item Case 2: Unlike Case 1, in this profile, the central absorption is not seen clearly, and it looks more similar to a third peak but with the violet and red peaks still distinguished on each side. %These kinds of H$\alpha$ emission lines are observed from 2014 to 2024. 
    The method used here was to locate the geometrical center of the line and the maximum emission at this center. The code then looks for the maximum intensities from each side of the center. The middle panel in Fig.~\ref{fig:cases} shows an example of this method. 
    \item Case 3: This case is similar to Case 2 but has a flat emission line, making it difficult for the code to distinguish between the two peaks. The intensity peaks were selected by visual inspection between the extremes from each side of the geometrical center of the line (see the right panel in Fig.~\ref{fig:cases}). We note that combining automatically computed values with those obtained by visual inspection can lead to inconsistencies and systematic biases. In particular, this could cause a larger scatter in our results. To diminish this possibility, we emphasized that higher peaks on each side were to be selected, and in cases where all peaks (when there are more than two) have similar intensities, the extremes of the peak emission were selected.
\end{itemize}

% \subsection{Radial velocity}
% For all spectral lines in Table~\ref{tab:lines}, we calculated the RV. Spectra from BeSS were corrected by barycentric velocity before measuring the RV. In the case of absorption lines, a Gaussian fit was performed, where the center of the fit was used to obtain the RV of the line. For the case of emission lines, we first convert the wavelength to velocity units by using the theoretical values from the National Institute of Standards and Technology (NIST) database\footnote{\url{https://physics.nist.gov/PhysRefData/ASD/lines_form.html}}. Then, we used the center of the full-width at half maximum (FWHM) of the emission line to measure the RV. We followed the Doppler formula to obtain the values.
 
\subsection{Errors for the violet to red peak intensity ratio and double-peak separation}
The spectra used in this work come from different instruments with different spectral resolutions. Table~\ref{tab:bess_combined} contain a list of all instruments and spectral resolutions for the available data.  To account for the fitting errors on the measurements of  V/R and DPS considering the different spectral resolutions, we used an open module from \texttt{Python}\footnote{\url{https://coronagraph.readthedocs.io/en/latest/index.html}}  to construct a wavelength grid by specifying the resolution (R=20000) and then rebinning the spectrum computing the mean within each bin, to the lower resolution in our sample (R=5800).  
%We chose a spectrum with R=20000, the highest resolution in our sample, to degrade it to the lowest resolution available in our data, R=5800. 
Then, we measured V/R and  DPS in both profiles and calculated the percentage error in Case 1 (and 2) for V/R and DPS, errors are 3$\%$ (and 2$\%$) and 12$\%$ (and 4$\%$), respectively. For case 3, we are considering the maximum error estimation found between both cases. In summary,  the uncertainties have accuracies of 3$\%$ and 12$\%$ for V/R and DPS, respectively.

%\begin{figure*}
%  \centreing
%  \includegraphics[scale=0.4]{Images/ERRORVRDPS.pdf}
%  \includegraphics[scale=0.4]{Images/ERRORVRDPS2.pdf}
%  \caption{Calculation of V/R with the case 1 in a spectra of R=20000 (Left) and R=6000 (Right) }
%  \label{fig:errorcase1vr}
%\end{figure*}

\subsection{Periods}
The high-cadence H$\alpha$ line emission data for $\rm \pi$~Aquarii presents a significant opportunity to analyze the periodicity of the V/R emission line variation. By employing the \texttt{astropy} library's implementation of the Lomb-Scargle periodogram \citep{Lomb1976,1982Scargle}, a sophisticated algorithm designed for the detection and characterization of periodic signals, we have successfully determined the period of these V/R shift variations. A key advantage of the Lomb-Scargle periodogram is its capacity to significantly reduce frequency aliasing, a phenomenon where higher frequency signals are misrepresented as lower frequencies due to inadequate sampling rates. This capability is crucial for accurately identifying periodic components, particularly those with frequencies exceeding the Nyquist frequency \citep{Nyquist1928}. In addition to the Lomb-Scargle periodogram, we use the date-compensated discrete Fourier transform \citep{1981Ferraz} that offers a robust method for analyzing unevenly spaced data. The date-compensated discrete Fourier transform is specifically adapted to handle irregular temporal sampling by employing the concept of function space projection to perform a Fourier transform.

\subsection{Determination of disk parameters}
To estimate the disk parameters of $\rm  \pi$~Aquarii, we used a grid of models created using the 3D Monte Carlo radiative transfer code \texttt{HDUST} \citep{Carciofi2006,Carciofi2008} to model $\rm  H\alpha$ lines and the spectral energy distribution (SED). This code solves the radiative and statistical equilibrium equations for hydrogen, considering non-LTE and different types of opacity and considering fast rotation for the central star; therefore, it assumes gravity darkening and oblateness effects. To model the circumstellar gaseous disk was used the steady-state power-law approximation of a viscous decretion disk (VDD) model \citep{Lee1991, Bjorkman1997}: the density distribution falls off as a power law radially ($r$) and a Gaussian distribution in hydrostatic equilibrium axisymmetric respect to the mid-plane of the disk in the vertical direction ($z$). In this approximation, the volumetric density distribution can be written as
\begin{equation}
    \rho(r,z) = \rho_{0} \left( \frac{r}{R_{eq}}  \right)^{-n} \exp \left(\frac{-z^{2}}{2H^{2}} \right),
\end{equation}
where $\rm \rho_{0}$ is the disk base density, $\rm R_{eq}$ is the equatorial radius of the central star, $\rm n$ is the volume density distribution index and $\rm H$ is the height scale that sets the flaring of Be stars and is given by
\begin{equation}
    H = H_{0} \left(\frac{r}{R_{eq}} \right)^{\beta}.
\end{equation}
For an isothermal disk, $\beta$=3/2 and the constant $H_{0}$ is defined as
\begin{equation}
    H_{0} = \left(\frac{2\, R_{eq}^{3} \,k\, T_{0}}{G\,M\, \mu_{0}\, m_{H}} \right)^{1/2},
\end{equation}
where $\rm M$ is the stellar mass, $\rm G$ is the gravitational constant, $m_{H}$ is the mass of a hydrogen atom, $\rm \rm k$ is the Boltzmann constant, $\rm \mu_{0}$ is the mean molecular weight of the gas and $\rm T_{0}$ is an isothermal temperature used only to fix the initial vertical structure of the disk. 

With a vertically integrated disk density $\rm \rho(r,z)$, we can obtain the surface mass density $\rm \Sigma(r)$ as follows:
\begin{equation}
   \Sigma(r)=\int_{-\infty }^{+\infty }\rho(r,z) dz.
\end{equation}

The relation between the volume and surface mass densities at the base of the disk $\Sigma_{0}$ is given by

\begin{equation}
\rho_{0}=\Sigma_{0}\sqrt{\frac{GM}{2\pi {c_{s}}^{2}{R_{eq}}^{3}}}.
\end{equation}

As input the code takes three disk parameters: the number density\footnote{The volumetric base density can be calculated as $\rm \rho_0=n_0 \mu m_H$ where $\rm m_H$ is the hydrogen mass and $\rm \mu$ is the mean molecular weight of the gas.} $\rm n_0$, the radial exponent $\rm n$, and the disk truncation radius $\rm R_{d}$. The isothermal temperature was fixed at $\rm T_0=0.72\ T_{eff}$ \citep{Correia2019}, and the mean molecular weight of hydrogen was $\rm \mu =\,0.6$. % and the gravity darkening exponent.  % $\beta_{GD}=\,0.25$ for all the models.

A source for the central star needed to be defined. %One option is the self-consistent rigid rotator model in the Roche approximation where three parameters as input are supplied: $\rm M$, $\rm W$, and $\rm t/t_{MS}$, where $t_{MS}$ is the age in the main sequence of the star. With these parameters as input, the photospheric parameters (R$_p$ and L) are calculated from the Geneva models \citep{Ekstrom2008,Georgy2013,Granada2013}, that computes the stellar evolution and structure of fast-rotating massive stars. The other option available is 
We selected a parametric rigid rotator providing the following parameters: $ \rm M$, $\rm R_p$, $ \rm R_{eq}/R_{p}$, $\rm L$, $\rm \log g$, $\rm V_{rot}$, and $\rm T_{eff}$. The stellar atmosphere models of \cite{Kurucz1994} are used for the stellar spectrum and solar metallicity ($\rm Z=0.014$) was used. The limb darkening models from \cite{Claret2000} are adopted to outline the photospheric specific intensity as a function of direction. 

The \texttt{HDUST} code was used to create a grid of models called BeAtlas \citep{Rubio2023}. Our version of this grid contains a total of 13800 models, which include Balmer lines and SED in the range between 0.10 $\rm \mu m$ and 2.45 $\rm \mu m$. The age adopted for the grid models is fixed and has been parameterized using the fraction of hydrogen in the core $\rm X_{c}$ = 0.3, which corresponds to the end of the main sequence, where most Be stars are concentrated. 

The model grid adopts the normalized value (between 0 and 1) for the base surface density of the disk $\rm \Sigma_{0}$ during modeling for two reasons: the lower and upper limits are separated by a factor of $\rm \sim$ 10$^{2}$, which makes it difficult to explore the parameter space and the fact that more massive stars have much denser discs when compared to less massive stars \citep{Rimulo2018}.

The parameters contained in the model grid are 11 values of masses ($\rm M$), which correspond to spectral types between B0.5 and B8 \citep{Townsend2004},  5 values of oblateness ($\rm R\rm_{eq}$/\rm R$\rm_{p}$) which corresponds to different rotation rates of the star, the normalized base surface density of the disk ($\rm \Sigma_{0}$), 4 values of volume density distribution index ($\rm n$) which includes a realistic scenario of the disk, with values that indicate a dissipation ($\rm n=3$), stationary ($\rm n=3.5$) and growth ($\rm n = 4.0$ and $\rm n = 4.5$) of the disk \citep{Haubois2012}, and 10 values of the cosine of the inclination angle of the disk concerning the observer’s line of sight, $\rm cos(i)$, the value equal to 1 corresponds a pole-on view while the value equal to 0 an edge-on view. A range of each parameter is shown in Table~\ref{tab:beatlasparams}.

 \begin{table}[H]
        \caption[]{Parameters in the \texttt{BeAtlas} grid.}
    \centering\label{tab:beatlasparams}
        \begin{tabular}{ll} 
                \hline\hline
                \noalign{\smallskip}
                $\rm M [M_{\odot}]$ & 14.6, 12.5, 10.8, 9.6, 8.6, 7.7, 6.4, 5.5, \\
                            &  4.8, 4.2, 3.8 \\
                            
                Oblateness $\rm (R_{eq}/R_{p})$ & 1.1, 1.2, 1.3, 1.4, 1.45   \\
  
                $\rm n$  & 3.0, 3.5, 4.0, 4.5 \\
                cos(i) & 1.0, 0.89, 0.78, 0.67, 0.55, 0.44, \\
                   &  0.33, 0.22, 0.11, 0. \\
                $\rm \Sigma_{0}$ ($\rm g\,cm^{-2}$)$^{*}$  & 0.02, 0.05, 0.12, 0.28, 0.68, 1.65, 4.00 \\
        \hline
        \end{tabular}
 \tablefoot{($^{*}$) The star was modeled using the normalized surface density at the base of the disk (value between 0 and 1).  The true value is obtained after considering the star's mass and the normalized value.}
\end{table}

In addition, through the modeled parameters, the use of Geneva stellar evolution models \citep{Georgy2013,Granada2013} we can estimate other stellar parameters such as luminosity ($\rm \log$ L) and polar radius ($ \rm R_p$). The effective temperature at the pole ($\rm T_{p}$) using Eq.(4.21) from \citet{Cranmer1996}, polar surface gravity ($ \rm \log$ g$_p$) from its relationship with the mass and polar radius, the fraction of critical angular velocity ($\rm \omega$) using Eq.(9) from \citet{Ekstrom2008} and gravity darkening exponent ($ \rm \beta_{GD}$) through the model presented by \citet{EspinosaLara2011}, where the exponent depends on the rotation rate $\rm W$.

To determine the model that best fits our observations, we used the Bayesian code \texttt{EMCEE}\footnote{Available at: \url{https://github.com/dfm/emcee}} \citep{Emcee2019}, which uses a variation of the Markov chain Monte Carlo (MCMC) algorithm from \citet{Goodman2010}. This code is an ensemble sampler. It uses the concept of “walkers” to explore the multidimensional parameter space. Numerous of these walkers, defined by the user, are started together and, throughout the iterations, start exploring the parameter space independently (but connected), creating large chains that tend to converge on one value or more (in cases where a bi-modal is observed). As a result, we obtain the probability density functions (PDFs) for each parameter of our grid, which we use to estimate the errors and observe possible correlations between them. Among the advantages of using EMCEE code, we can mention $\rm i$) high efficiency when modeling multidimensional parameter space and $\rm ii$) parallelization of operations, reducing computation time. 

According to Bayesian formalism, we must define the likelihood function and a priori distribution. We used the likelihood function, $ \rm ln\,P(D|\theta,I)$, of the$  \rm\chi^{2}$-distribution type given by
\begin{equation}
ln\,P(D|\theta,I)\propto -\tfrac{1}{2}\chi^{2} ,
\end{equation}
where $ \rm \chi^{2}$ is defined as 
\begin{equation}
\label{eq:chisquare}
 \chi^{2} = \sum_{i=1}^{N} \frac{({F_{obs,i}-F_{mod,i}})^{2}}{{\sigma_{obs,i}}^{2}}.
\end{equation}
The logarithm of the prior distribution, $ln\,P(\theta,I)$, corresponds to the knowledge available about each of the parameters in our grid or any other relevant information that helps the modeling process. To follow the same range of parameters used in the stellar parameters determination, we have adopted a non-informative prior distribution for the mass range between 10 and 14.6 solar masses and inclination of the disk between 50 and 75 degrees, which correspond in terms of cos(i) to a range between 0.25 and 0.65 according to the information available in the literature (see the introduction). For the other 3 parameters, we also adopt a uniform prior distribution, but that covers all the values available in our grid. Another parameter we adopted as a prior distribution was the $\rm v\,$sin$\,i$; in this case, we use a Gaussian distribution (similar to Eq. \ref{eq:chisquare}) with mean 227 km\,s$^{-1}$ and standard deviation 78 km\,s$^{-1}$ from \citet{Solar2022}. For SED modeling, we include two other parameters: E(B-V) and the parallax, both using a Gaussian distribution according to the values obtained by \citep{Fitzpatrick1999} and Gaia DR3 \citep{Gaia2023}, respectively.

We modeled a total of five H$\alpha$ lines obtained at different years (i.e., 2001, 2011, 2014, 2018, and 2022) and SED using photometry data available in the Virtual Observatory VOSA \citep{Bayo2008} from missions and observers in ultraviolet, including the IUE mission from \citep{Benvenuti1983}, visible; UBV photometry from \citep{Mermilliod2006}; Stroemgren V filter from \citep{Paunzen2015}; Tycho-2 Catalogue from \citep{Hog2000}; Hipparcos catalog \citep{Hipparcos1997}; and Gaia DR3 \citep{Gaia2016}, and in the near-infrared the 2MASS catalog from \citep{Skrutskie2006}. The selected range of 0.14 to 2.4 $\mu m$, corresponds to the spectral coverage of our grid.

A total of 150 walkers were defined for all the modeling. To check the convergence of the walkers, we followed the recommendations given in the EMCEE \texttt{documentation},\footnote{Available at: \url{https://emcee.readthedocs.io/en/stable/}} where the number of iterations reaches the value corresponding to 50 times the correlation time and shows a variation of less than 1\% in subsequent iterations.

\section{Results}
\subsection{Stellar parameters obtained}
\begin{figure}
     \centering
     \includegraphics[scale=0.35]{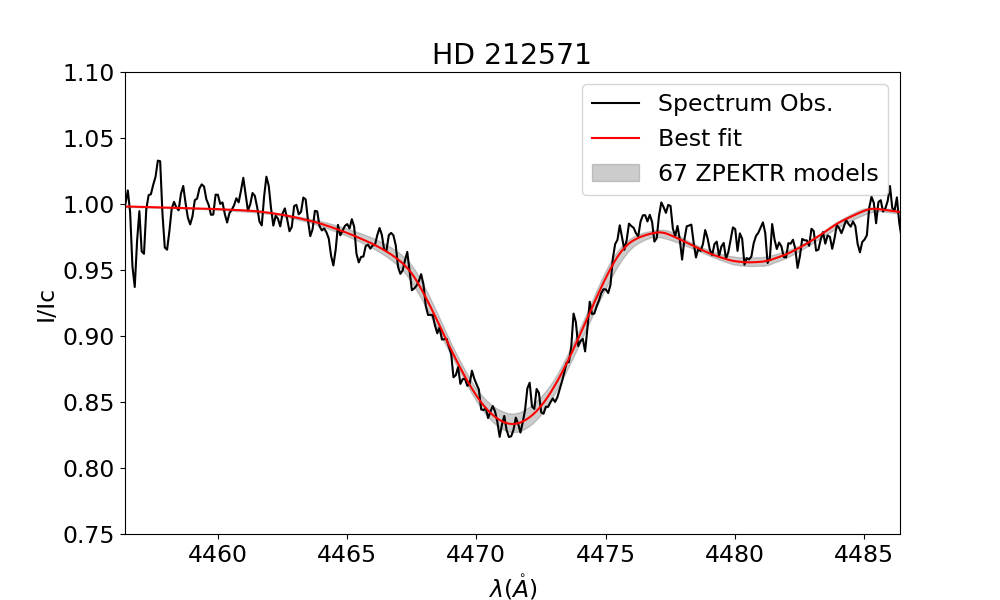}  
     \caption[]{Observed spectrum of the photospheric line HeI 4471 \AA \,(in black). The red line shows the best fit of the ZPEKTR models, and the gray band represents the flux values of the 67 best models.}\label{fig:HD212571a}
 \end{figure}

We created 11294 models with \texttt{ZPEKTR}, where the best models are 67. We obtained the average and standard deviation from the output parameters. The best-fit models are shown in Fig~\ref{fig:HD212571a}, where the gray band indicates the range of the best 67 models. The stellar parameters derived from these models are presented in Table~\ref{tab:HD212571}. The uncertainties were calculated as the standard deviations from the 67 models compared to the best-fit model.

Since $\rm W$ is not provided by \texttt{ZPEKTR}, it can be calculated using Eq. (11) from \cite{Rivinius2013}. Applying standard error propagation theory, we obtained a final value of $\rm W = 0.55 \pm 0.01$. 

 \begin{table}
\caption[]{Stellar parameters obtained with \texttt{ZPEKTR}.}
\label{tab:HD212571}
\centering
\begin{tabular}{cc} 
\hline\hline
                \noalign{\smallskip}
            $<\rm T_{eff}>$ [K] & 23849 $\!\pm$ 1138 \\
                $<\log g>$  & 4.03           $\!\pm$ 0.10   \\
                $\rm R$ \ [R$_{\odot}$] & 5.33\!$\pm$ 0.87 \\
             M [M$_{\odot}$] & 11.0      $\!\pm$ 1.9   \\
                $\log$ L \ [$ \rm L_{\odot}$] & 3.92 $\!\pm$ 0.08\\
            R$\rm _p$ \ [R$_{\odot}$]     & 5.00 $\!\pm$ 0.81 \\
            R$\rm _{eq}$ \ [R$_{\odot}$] & 5.75 $\!\pm$  0.99 \\
                % $\rm R_{eq}/R_p$ & 1.275  & $\pm$   0.088 \\
                $\rm  T_p$ [K] & 25000         $\!\pm$ 1174  \\
            $\rm  T_{eq}$ [K] & 22383      $\!\pm$ 1357  \\
            $\log g \rm _p$  & 4.08        $\!\pm$ 0.10   \\
            $\log g \rm _{eq}$  & 3.96     $\!\pm$ 0.11   \\
                $\rm  v \sin i$ [km/s]  & 271  $\!\pm$ 13 \\
                $\rm i[^{\circ}]$  & 55        $\!\pm$ 8  \\
                $\gamma$  & 0.63               $\!\pm$ 0.06  \\
                $\omega$  & 0.82               $\!\pm$ 0.05  \\
        \hline
        \end{tabular}
\end{table}

\subsection{Obtained disk parameters}
The observed profiles of the H$\alpha$ lines are normalized in flux. For this reason, we did not include parallax and $ \rm E(B-V)$ as additional parameters in the modeling since, for an isolated line, we can disregard any significant changes in its profile. The profiles were interpolated to the same resolution as the models contained in the grid. In addition, we used the velocity space, centered on the H$\alpha$ line, with a range between -1100 and +1100 $  \rm km\,s^{-1}$, sufficient to show the line profile and its wings. 

Figure ~\ref{fig:corner2001} (upper right inset) shows the results obtained for the spectrum observed in 2001. We can note that the PDFs are defined for mass, oblateness, $ \rm cos\,i$, and normalized $ \rm \Sigma _{0}$, the latter with a smaller uncertainty than the others. The $\rm n$ index showed an irregular distribution, almost flat. The line profile observed was well-fitted, both in the central region and on the wings. Using the Spearman coefficients,\footnote{Available as a function from Python's scipy package \url{https://scipy.org/}} we can also obtain information about the correlation (positive values) or anticorrelation (negative values) between each pair of parameters. The higher the value and the intensity of the color, the stronger the correlation between the parameters. We can see strong correlations between the normalized base density $\rm \Sigma_{0}$ and the $ \rm n$ index.

Figure~\ref{fig:corner2011} shows the results obtained for the spectrum observed in 2011. On this epoch, we can observe an intense double-peak profile with $\rm V/R \sim 1$. The fit obtained by the model was well-defined, as were the PDFs for each parameter. Similar to the results of 2001, an intense correlation between the disk parameters can also be observed.

Figure~\ref{fig:corner2014} shows the results obtained for the spectrum observed in 2014, where we can note a bimodality in the PDFs of three parameters, with the presence of very pronounced peaks, except for $\rm \Sigma_{0}$ and $\rm cos\,i$, this can indicate that both solutions are probably valid. The correlation/anticorrelation between the parameters is complex, probably due to bimodality. The modeled line profile reasonably agrees with the observed profile, especially on the left wing, where contamination appears to increase the flux in this region.

Figures~\ref{fig:corner2018} and \ref{fig:corner2022} show the results for the spectra observed in 2018 and 2022, respectively. In both cases, the observed line profiles were not well-fitted. This is due to the complexity of these lines, which have a width greater than 500 $ \rm km\,s^{-1}$ and asymmetrical $ \rm V/R$ values. Our version of the grid has limitations, including the lack of models with asymmetric line profiles caused by inhomogeneities in the disk and the non-inclusion of physical processes, such as electronic scattering, which would broaden the line profiles. Although the PDFs are well-defined, with pronounced peaks, the values obtained for the parameters are not reliable. 

Figure~\ref{fig:cornersed} shows the results of the SED modeling. Although the photometric data were obtained at different epochs and could be affected by the variability caused by the disk formation/dissipation process, we obtained a good fit. One possible explanation may be related to the coverage of the grid between the ultraviolet and the near-infrared since, in this region, the star's parameters are better defined than the parameters of the disk. Therefore, they are less affected by photometric variations coming from the disk. This is evident from the poorly defined PDFs of the disk parameters. Both E(B-V) and parallax, included in the SED modeling, showed well-defined PDFs. Table~\ref{tab:Stellar_emcee} and \ref{tab:Disk_emcee} contain stellar and disk parameters, respectively, obtained as a result of the modeling of the H$\alpha$ line observed on different dates and the SED. 

\begin{table*}
\centering
\tiny
\caption{Stellar parameters obtained through the application of the EMCEE code.}
\label{tab:Stellar_emcee}
 \renewcommand{\arraystretch}{1.3}
\begin{tabular}{lllllllllllllllll}
Line & Date & M & $\rm R_{eq}/R_p$ & R$\rm _p$ & $\log$ L$\rm_{p}$ & $\log$ g$\rm_{p}$ & $\rm  v$ & $\rm  T_{p}$ & $\beta_{GD}$&$\rm  E(B-V)$&$\rm plx$ \\
     &  & [M$_{\odot}$] &  & [R$_{\odot}$]  & [$ \rm L_{\odot}$]  & & [km/s] & [K] &  &\\
\hline
\hline
H$\alpha$ & 2001-12-22 & $12.23_{-1.12}^{+0.97}$ & $1.28_{-0.05}^{+0.05}$ & $6.67_{-0.46}^{+0.41}$ & $4.24_{-0.15}^{+0.12}$ & $3.88_{-0.01}^{+0.02}$ & $391.99_{-24.79}^{+19.76}$ & $26343_{-1379}^{+999}$ & $0.18_{-0.01}^{+0.01}$ \\
%H$\beta$  & 2001-12-22 & $14.36_{-0.35}^{+0.17}$ & $1.22_{-0.02}^{+0.03}$ & $7.48_{-0.13}^{+0.07}$ & $4.47_{-0.03}^{+0.02}$ & $3.85_{-0}^{+0.01}$ & $359_{-12.64}^{+18.75}$ & $27916_{-284}^{+181}$ & $0.19_{-0.01}^{+0}$\\
%H$\gamma$ & 2001-12-22 & $14.4_{-0.43}^{+0.13}$ & $1.19_{-0.02}^{+0.01}$ & $7.52_{-0.16}^{+0.04}$ & $4.47_{-0.04}^{+0.01}$ & $3.85_{-0}^{+0.01}$ & $343.9_{-12.99}^{+10.58}$ & $27993_{-392}^{+113}$ & $0.2_{-0}^{+0}$\\
%H$\delta$ & 2001-12-22 & $14.52_{-0.13}^{+0.06}$ & $1.14_{-0.02}^{+0.02}$ & $7.6_{-0.06}^{+0.03}$ & $4.49_{-0.02}^{+0.01}$ & $3.84_{-0}^{+0}$ & $299.23_{-19.61}^{+22.28}$ & $28089_{-120}^{+80}$ & $0.21_{-0.01}^{+0}$\\
H$\alpha$ & 2011-10-03 & $14.59_{-0.03}^{+0.01}$ & $1.1_{-0.01}^{+0.01}$ & $7.67_{-0.01}^{+0.01}$ & $4.5_{-0.01}^{+0.01}$ & $3.83_{-0.0}^{+0.01}$ & $258.54_{-1.35}^{+2.67}$ & $28071_{-26}^{+14}$ & $0.22_{-0.01}^{+0.01}$ \\
%H$\beta$  & 2011-10-03 & $14.54_{-0.1}^{+0.05}$ & $1.11_{-0}^{+0.01}$ & $7.64_{-0.03}^{+0.03}$ & $4.49_{-0.01}^{+0.01}$ & $3.83_{-0}^{+0}$ & $264.27_{-5.7}^{+9.03}$ & $28043_{-92}^{+51}$ & $0.22_{-0.01}^{+0.01}$\\
%H$\gamma$ & 2011-10-03 & $14.52_{-0.34}^{+0.07}$ & $1.4_{-0.25}^{+0.02}$ & $7.53_{-0.13}^{+0.06}$ & $4.48_{-0.03}^{+0.01}$ & $3.85_{-0.01}^{+0.01}$ & $459.34_{-148.32}^{+5.66}$ & $30495_{-2356}^{+255}$ & $0.16_{-0}^{+0.05}$ \\
H$\alpha$ & 2014-10-29 & $14.49_{-0.3}^{+0.09}$ & $1.19_{-0.07}^{+0.01}$ & $7.54_{-0.06}^{+0.04}$ & $4.48_{-0.02}^{+0.01}$ & $3.84_{-0.01}^{+0.01}$ & $341.34_{-62.57}^{+7.2}$ & $28081_{-291}^{+94}$ & $0.20_{-0.01}^{+0.02}$ \\
%H$\beta$  & 2014-10-29 & $13.95_{-0.09}^{+0.25}$ & $1.15_{-0.02}^{+0.03}$ & $7.35_{-0.03}^{+0.11}$ & $4.43_{-0.01}^{+0.03}$ & $3.85_{-0}^{+0}$ & $312.8_{-17.36}^{+26.69}$ & $27592_{-92}^{+243}$ & $0.21_{-0.01}^{+0}$\\
%H$\gamma$ & 2014-10-29 & $14.54_{-0.2}^{+0.05}$ & $1.4_{-0.28}^{+0.02}$ & $7.54_{-0.03}^{+0.04}$ & $4.48_{-0.01}^{+0}$ & $3.85_{-0.01}^{+0}$ & $460.73_{-175.29}^{+6.5}$ & $30550_{-2577}^{+205}$ & $0.16_{-0}^{+0.06}$\\
H$\alpha$ & 2018-10-09 &  $14.6_{-0.01}^{+0.01}$ & $1.1_{-0.01}^{+0.01}$ & $7.68_{-0.01}^{+0.01}$ & $4.5_{-0.01}^{+0..01}$ & $3.83_{-0.01}^{+0.01}$ & $257.53_{-0.54}^{+1.09}$ & $28082_{-5}^{+2}$ & $0.22_{-0.01}^{+0.01}$ \\
%H$\beta$  & 2018-10-09 & $14.47_{-0.1}^{+0.08}$ & $1.11_{-0.01}^{+0.01}$ & $7.61_{-0.05}^{+0.04}$ & $4.48_{-0.01}^{+0.01}$ & $3.84_{-0}^{+0}$ & $271.02_{-7.92}^{+13.64}$ & $28006_{-77}^{+65}$ & $0.22_{-0.01}^{+0.01}$\\
%H$\gamma$ & 2018-10-09 & $14.58_{-0.04}^{+0.02}$ & $1.4_{-0.3}^{+0.05}$ & $7.55_{-0.01}^{+0.12}$ & $4.49_{-0}^{+0.01}$ & $3.85_{-0.01}^{+0}$ & $459.27_{-200.2}^{+17.67}$ & $30145_{-2081}^{+615}$ & $0.16_{-0.01}^{+0.06}$\\
%H$\alpha$ & 2022-09-03 & $14.6_{-0}^{+0}$ & $1.15_{-0}^{+0}$ & $7.62_{-0}^{+0}$ & $4.49_{-0}^{+0}$ & $3.84_{-0}^{+0}$ & $309.16_{-1.4}^{+1.38}$ & $28195_{-17}^{+12}$ & $0.21_{-0.01}^{+0.01}$\\
%H$\beta$  & 2022-09-03 & $14.58_{-0.03}^{+0.01}$ & $1.14_{-0.01}^{+0.01}$ & $7.62_{-0.01}^{+0.01}$ & $4.49_{-0}^{+0}$ & $3.84_{-0}^{+0}$ & $297.87_{-8.73}^{+8.51}$ & $28176_{-41}^{+35}$ & $0.21_{-0.01}^{+0.01}$\\
%H$\gamma$ & 2022-09-03 & $14.58_{-0.04}^{+0.02}$ & $1.11_{-0}^{+0.03}$ & $7.66_{-0.04}^{+0.02}$ & $4.5_{-0}^{+0}$ & $3.83_{-0}^{+0}$ & $263.93_{-5.8}^{+32.35}$ & $28081_{-40}^{+64}$ & $0.22_{-0.01}^{+0.01}$\\
%H$\delta$ & 2022-09-03 & $14.35_{-0.47}^{+0.2}$ & $1.13_{-0.02}^{+0.26}$ & $7.52_{-0.2}^{+0.1}$ & $4.47_{-0.05}^{+0.02}$ & $3.84_{-0.01}^{+0.01}$ & $284.01_{-23.44}^{+169.87}$ & $28057_{-396}^{+2163}$ & $0.21_{-0.06}^{+0.01}$\\
H$\alpha$ & 2022-11-25 & $13.19_{-0.02}^{+0.02}$ & $1.2_{-0.01}^{+0.01}$ & $7.02_{-0.01}^{+0.01}$ & $4.35_{-0.01}^{+0.01}$ & $3.87_{-0.01}^{+0.01}$ & $345.54_{-0.21}^{+0.12}$ & $26988_{-15}^{+13}$ & $0.20_{-0.01}^{+0.01}$ \\
%H$\beta$ & 2022-11-25 & $14.5_{-0.31}^{+0.08}$ & $1.11_{-0.01}^{+0.33}$ & $7.54_{-0.04}^{+0.13}$ & $4.48_{-0.02}^{+0.02}$ & $3.84_{-0.01}^{+0.01}$ & $270.4_{-10.57}^{+204.51}$ & $28065_{-284}^{+2085}$ & $0.22_{-0.07}^{+0}$\\
%H$\gamma$ & 2022-11-25 & $10.06_{-0.05}^{+0.19}$ & $1.16_{-0.05}^{+0.11}$ & $5.93_{-0.08}^{+0.06}$ & $3.97_{-0.01}^{+0.02}$ & $3.9_{-0.01}^{+0.01}$ & $295.86_{-49.1}^{+76.22}$ & $23735_{-133}^{+810}$ & $0.21_{-0.02}^{+0.01}$\\
SED       &            & $13.84_{-0.33}^{+0.23}$ & $1.11_{-0.01}^{+0.02}$ & $7.35_{-0.13}^{+0.1}$ & $4.42_{-0.03}^{+0.02}$ & $3.85_{-0.01}^{+0.01}$ & $267.96_{-8.74}^{+20.37}$ & $27481_{-295}^{+159}$ & $0.22_{-0.01}^{+0.01}$ &$0.15_{-0.01}^{+0.01}$ &$2.58_{-0.02}^{+0.05}$\\
\hline
\noalign{\smallskip}
\end{tabular}
%\end{sidewaystable}
\end{table*}

\begin{table*}
\centering
\tiny
\caption{Disk parameters obtained through the application of the EMCEE code.}
\label{tab:Disk_emcee}
 \renewcommand{\arraystretch}{1.3}
\begin{tabular}{lllllllllll}
Line & Date         & $\rm \Sigma_{0}$ & $\omega$ & $\rm W$ & $\rm n$ & $\rm cos(i)$ & $\rm i(^{\circ})$  \\
     & (YYYY-MM-DD) &                  &          &         &         &              &      \\
\hline
\hline
H$\alpha$ & 2001-12-22 & $0.01_{-0.01}^{+0.01}$ & $0.95_{-0.03}^{+0.02}$ & $0.75_{-0.06}^{+0.06}$ & $3.93_{-0.45}^{+0.38}$ & $0.50_{-0.10}^{+0.08}$ & $59.9_{-4.96}^{+6.81}$ \\
%H$\beta$  & 2001-12-22 & $0.03_{-0.01}^{+0.01}$ & $0.9_{-0.02}^{+0.02}$ & $0.66_{-0.03}^{+0.04}$ & $3.71_{-0.29}^{+0.47}$ & $0.44_{-0.07}^{+0.07}$ & $63.14_{-4.68}^{+4.56}$   \\
%H$\gamma$ & 2001-12-22 & $0.04_{-0}^{+0.25}$ & $0.88_{-0.02}^{+0.02}$ & $0.62_{-0.03}^{+0.02}$ & $3.61_{-0.31}^{+0.33}$ & $0.47_{-0.07}^{+0.07}$ & $61.1_{-3.97}^{+4.6}$ \\
%H$\delta$ & 2001-12-22 & $0.04_{-0.01}^{+0.01}$ & $0.8_{-0.04}^{+0.04}$ & $0.53_{-0.04}^{+0.04}$ & $3.98_{-0.28}^{+0.35}$ & $0.33_{-0.07}^{+0.09}$ & $71.26_{-5.69}^{+4.2}$ \\
H$\alpha$ & 2011-10-03 & $0.28_{-0.01}^{+0.01}$ & $0.72_{-0.01}^{+0.01}$ & $0.45_{-0.01}^{+0.01}$ & $3.20_{-0.01}^{+0.01}$ & $0.34_{-0.01}^{+0.01}$ & $70.42_{-0.14}^{+0.14}$ \\
%H$\beta$  & 2011-10-03 & $0.25_{-0.01}^{+0.03}$ & $0.73_{-0.01}^{+0.02}$ & $0.46_{-0.01}^{+0.02}$ & $3.06_{-0.04}^{+0.07}$ & $0.34_{-0.01}^{+0.01}$ & $70.01_{-0.81}^{+0.57}$  \\
%H$\gamma$ & 2011-10-03 & $0.05_{-0.01}^{+0.36}$ & $0.99_{-0.17}^{+0}$ & $0.9_{-0.34}^{+0.02}$ & $3.35_{-0.27}^{+0.13}$ & $0.43_{-0.05}^{+0.07}$ & $64.43_{-5.2}^{+2.38}$ \\
H$\alpha$ & 2014-10-29 & $0.10_{-0.01}^{+0.05}$ & $0.87_{-0.11}^{+0.01}$ & $0.61_{-0.12}^{+0.02}$ & $3.16_{-0.02}^{+0.21}$ & $0.53_{-0.01}^{+0.05}$ & $58.25_{-3.50}^{+0.52}$ \\
%H$\beta$  & 2014-10-29 & $0.18_{-0.01}^{+0.03}$ & $0.83_{-0.03}^{+0.04}$ & $0.56_{-0.03}^{+0.05}$ & $3.32_{-0.04}^{+0.07}$ & $0.59_{-0.01}^{+0.01}$ & $53.71_{-0.47}^{+0.72}$ \\
%H$\gamma$ & 2014-10-29 & $0.08_{-0.02}^{+0.34}$ & $0.99_{-0.22}^{+0}$ & $0.9_{-0.4}^{+0.02}$ & $3.29_{-0.12}^{+0.11}$ & $0.54_{-0.07}^{+0.05}$ & $57.28_{-3.21}^{+4.37}$ \\
H$\alpha$ & 2018-10-09 &  $0.65_{-0.01}^{+0.01}$ & $0.71_{-0.01}^{+0.01}$ & $0.45_{-0.01}^{+0.01}$ & $3.00_{-0.01}^{+0.01}$ & $0.44_{-0.01}^{+0.01}$ & $63.59_{-0.03}^{+0.03}$ \\
%H$\beta$  & 2018-10-09 & $0.47_{-0.02}^{+0.02}$ & $0.74_{-0.02}^{+0.03}$ & $0.47_{-0.02}^{+0.03}$ & $3.01_{-0.01}^{+0.02}$ & $0.56_{-0.01}^{+0.01}$ & $55.66_{-0.46}^{+0.47}$  \\
%H$\gamma$ & 2018-10-09 & $0.48_{-0.1}^{+0.73}$ & $0.99_{-0.27}^{+0.01}$ & $0.9_{-0.44}^{+0.05}$ & $3.13_{-0.06}^{+0.17}$ & $0.5_{-0.03}^{+0.07}$ & $59.8_{-4.17}^{+1.94}$ \\
%H$\alpha$ & 2022-09-03 & $0.68_{-0}^{+0}$ & $0.82_{-0}^{+0}$ & $0.55_{-0}^{+0}$ & $3_{-0}^{+0}$ & $0.28_{-0}^{+0}$ & $73.9_{-0.09}^{+0.08}$   \\
%H$\beta$  & 2022-09-03 & $0.68_{-0.01}^{+0.01}$ & $0.8_{-0.02}^{+0.02}$ & $0.53_{-0.02}^{+0.02}$ & $3_{-0}^{+0}$ & $0.35_{-0.01}^{+0.01}$ & $69.74_{-0.58}^{+0.53}$ \\
%H$\gamma$ & 2022-09-03 & $2.33_{-1.03}^{+0.21}$ & $0.73_{-0.01}^{+0.07}$ & $0.46_{-0.01}^{+0.06}$ & $3.1_{-0.09}^{+0.04}$ & $0.6_{-0.02}^{+0.03}$ & $52.92_{-2.4}^{+1.17}$ \\
%H$\delta$ & 2022-09-03 & $3.12_{-2.9}^{+0.48}$ & $0.77_{-0.05}^{+0.22}$ & $0.5_{-0.04}^{+0.39}$ & $3.27_{-0.07}^{+0.42}$ & $0.64_{-0.05}^{+0.02}$ & $50.3_{-1.26}^{+2.4}$  \\
H$\alpha$ & 2022-11-25 & $0.68_{-0.01}^{+0.01}$ & $0.88_{-0.01}^{+0.01}$ & $0.63_{-0.01}^{+0.01}$ & $3.00_{-0.01}^{+0.01}$ & $0.22_{-0.01}^{+0.01}$ & $77.01_{-0.02}^{+0.01}$ \\
%H$\beta$ & 2022-11-25  & $0.82_{-0.35}^{+0.04}$ & $0.74_{-0.02}^{+0.25}$ & $0.47_{-0.02}^{+0.46}$ & $3.04_{-0.02}^{+0.02}$ & $0.47_{-0.06}^{+0}$ & $62.25_{-0.26}^{+4.46}$ \\
%H$\gamma$ & 2022-11-25 & $3.07_{-0.4}^{+0.36}$ & $0.82_{-0.1}^{+0.11}$ & $0.56_{-0.1}^{+0.16}$ & $3.36_{-0.03}^{+0.03}$ & $0.66_{-0.01}^{+0}$ & $48.64_{-0.22}^{+0.78}$ \\
SED       &           & $0.01_{-0.01}^{+0.01}$ & $0.74_{-0.02}^{+0.04}$ & $0.47_{-0.02}^{+0.04}$ & $4.04_{-0.51}^{+0.33}$ & $0.57_{-0.15}^{+0.07}$ & $54.28_{-4.22}^{+8.71}$ \\
\hline
\noalign{\smallskip}
\end{tabular}
\end{table*}

\subsection{Measurements obtained over H$\alpha$ emission line: EW, V/R and DPS.}
%The result of each measurement is presented in the following subsections.

\subsubsection{Equivalent width variability}
\begin{figure*}
\begin{center}
\includegraphics[scale=0.4]{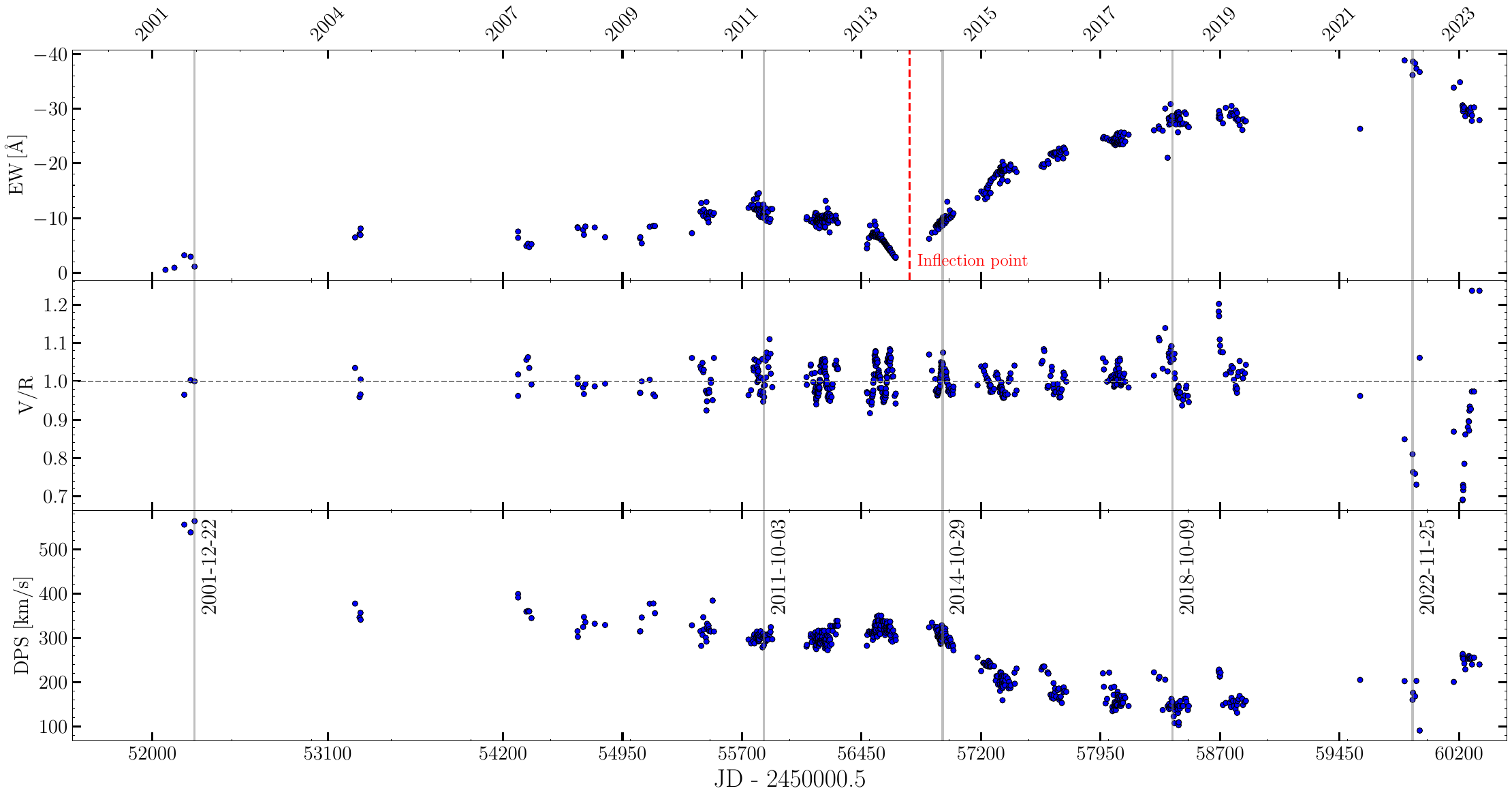}
\caption{Equivalent width, V/R, and DPS of H$\alpha$ plotted against time in MJD. Gray vertical lines denote five specific dates with significant changes in EW, which are discussed in the text. A horizontal line at $\rm V/R = 1$ is included to highlight the asymmetry of H$\alpha$.} 
\label{fig:haobserv}
\end{center}
\end{figure*}

Figure~\ref{fig:haobserv} shows the evolution of H$\alpha$ EWs, V/R, and DPS from 2001 to 2024. . The EW values of H$\alpha$ were obtained for all observation epochs and corrected for photospheric absorption using synthetic spectra from \texttt{ZPEKTR}. These values are available in an \href{https://docs.google.com/spreadsheets/d/1jDAFGvOOEU8OPzYMJQ7btGLXoisBngnSnmS6OdfEo8A/edit#gid=764686422}{online spreadsheet}. Thus, the H$\alpha$ EWs values represent `pure' emission, and therefore, the growth of the EW curve will indicate the disk growth, as well as the decrease of the EW curve, which will indicate a disk dissipation phase. We note to the reader that all results in this work do not consider the effects of the continuum in the calculated EW. This is because most spectra are already normalized to the continuum in the BeSS and BeSOS databases. Possible effects are described in the discussion section. The sharp increase in recent years may be related to episodes of enhanced mass-loss rates or outflows. Gray vertical lines in Fig.~\ref{fig:haobserv} indicate five significant epochs where noticeable changes in EW occur, providing reference points to help explain the time variability: MJD\footnote{Modified Julian Dates} 52265 (December 22, 2001), 55837 (October 03, 2011), 56959 (October 29, 2014), 58400 (October 09, 2018), and 59908 (November 25, 2022).

Due to the lack of earlier data in this study, no conclusions can be drawn about the growth or dissipation of the disk before 2001. However, \cite{Miroshnichenko2002} reported the lowest brightness level observed in $\pi$~Aquarii in 1998, along with weak H$\alpha$ emission, which they described as a `quasi-normal star' phase between 1996 and 2001. Between December 2001 and October 2011, the EW grew slowly and fluctuated around -10 \AA\,.  Then, from October 2011, the EW decreased, reaching a minimum documented value close to -2 \AA\, and making an inflection point (in October 2014 and marked with a red dashed vertical line) in the behavior curve where it begins to increase its value steadily over time until it reaches a value of -28.7 \AA\, in October 2018 (at the fourth gray vertical line). Subsequently, the EW fluctuates around this value until it increases drastically, reaching a value of -36.1\AA \ in November 2022 (at the fifth gray vertical line), which is more than 18 times the value at the inflection point. This suggests that the growth of the circumstellar envelope has continued without interruption since October 2014; an outflow or enhanced mass-loss rate occurred in recent years (November 2022). After this, the EW started to decrease rapidly.

Table~\ref{tab:comparisonewbalmer} displays the EW ratio between the Balmer lines and H$\alpha$ for the five selected observation dates. The results show that the ratio amplitude increases toward the H$\gamma$ line, with similar EW behavior observed for all the Balmer lines.

\begin{table*}
\centering
\caption{EW and V/R ratios of Balmer lines for five selected dates.}
\label{tab:comparisonewbalmer}
\begin{tabular}{cccccc}  
        Dates [MJD] & 52265 & 55837 & 56959 & 58400 & 59908 \\ 
                    & Dec 22, 2001 & Oct 03, 2011 & Oct 29, 2014 & Oct 09, 2018 & Nov 25, 2022 \\ 
        \hline
        \hline
        EW H$\alpha$\ [\AA] & -1.15 & -11.4 & -8.7 & -28.7 & -36.1 \\ 
        EW H$\beta$/H$\alpha$ & 1.72 & 0.29 & 0.34 & 0.16 & 0.19 \\ 
        EW H$\gamma$/H$\alpha$ & 1.79 & 0.24 & 0.31 & 0.11 & 0.09 \\ 
        V/R H$\alpha$ & 1.00 & 0.99 & 1.05 & 1.07 & 0.81 \\ 
        \hline
\end{tabular}
\tablefoot{
The five observation dates are marked with a gray vertical line in Fig.~\ref{fig:haobserv}. The first row contains the EW values of H$\alpha$. The other rows show the ratio between the other Balmer lines and EW H$\alpha$.
}\\
\end{table*}

\subsubsection{Violet to red peak intensity ratio variability}
The V/R  presents periodical fluctuations with intensity values between 0.69 and 1.24 over time. Despite this, these variations have some differences depending on the epochs. For example, between October 2011 and 2014 (second and third gray vertical lines), in the zones with a high concentration of observations, the values describe an M-type shape (typical of a cyclical variation) around one. We note that before the EW plot's inflection point (October 2014), the V/R intensity ratio reached a higher and deeper value than the previous months. After the inflection point, the lowest values are less deep, and the highest values are more intense, becoming asymmetric and biased toward the violet peak. Since October 2014 (third gray vertical line \footnote{All vertical lines are referred to Fig. \ref{fig:haobserv}}), the V/R ratio has become confusing as the disk grows. This can be a result of the precession due to the one-armed density wave presented in the disk. These values could also result from losing the symmetrical shape of the H$\alpha$ emission line; as can be seen in the DPS curve, values are lower, indicating a more extended region of the emitting area.  

\begin{figure}
\begin{center}
\includegraphics[width=\columnwidth]{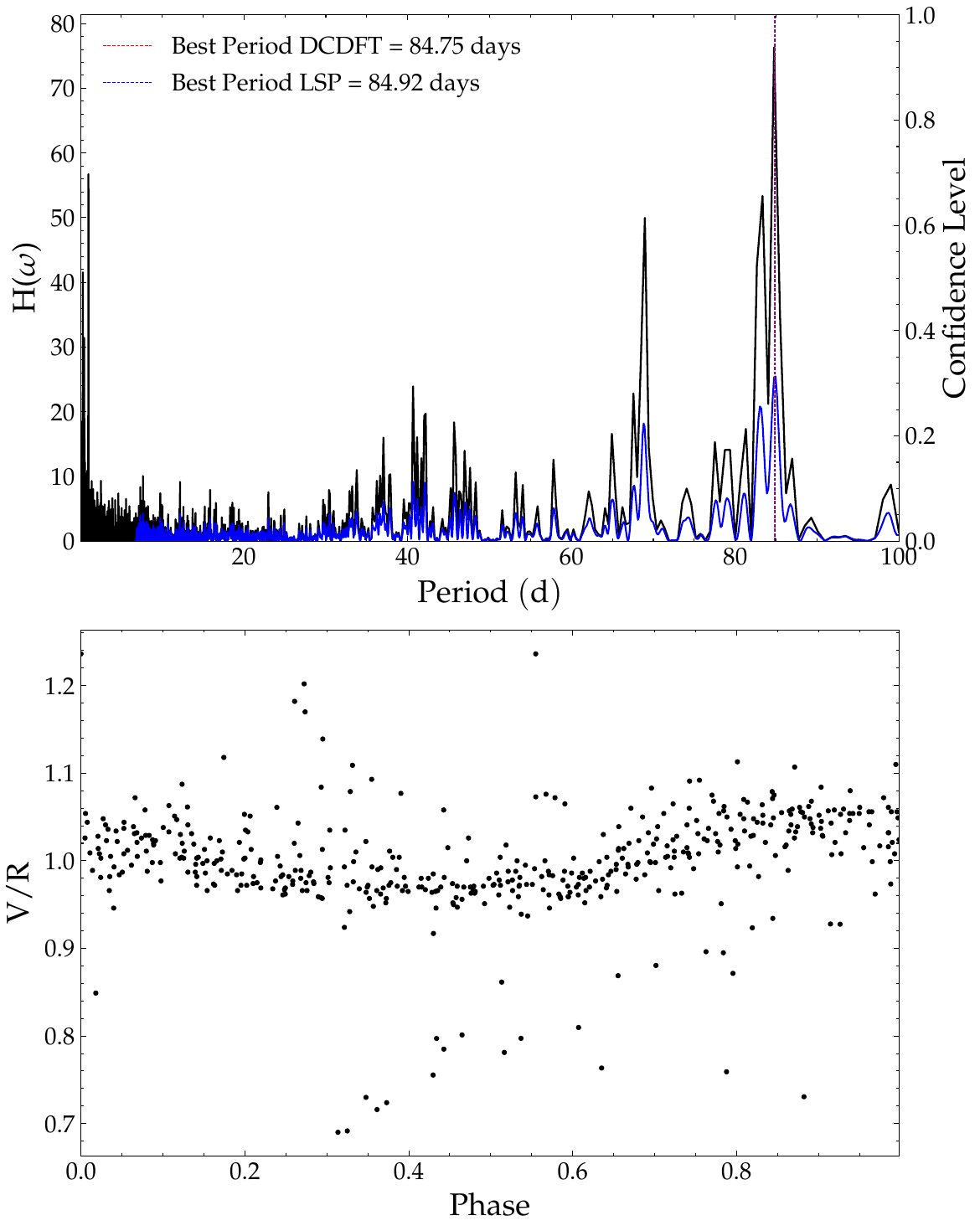}
\caption{\textit{Top panel:} Power spectrum for the V/R presented in units of periods. The date-compensated discrete Fourier transform periodogram is shown as a black solid line, while the Lomb-Scargle periodogram is represented as a blue solid line. Both periodograms have been normalized by their respective means to enhance visibility. \textit{Bottom panel:} Phase of the V/R computed with the period of 84.75 days. }
\label{fig:periodvr}
\end{center}
\end{figure}

The Lomb-Scargle periodogram shown in Fig.~\ref{fig:periodvr} has been computed for the V/R  of H$\alpha$. The left panel shows the power spectrum. The maximum amplitude corresponds to 81.75 days for the date-compensated discrete Fourier transform periodogram and 81.94 days for the Lomb-Scargle periodogram. The other maximum frequency corresponds to 69 days. The right panel displays the phase of the V/R  from 0 to 1, calculated with the best period of 84.75 days, which exhibits a sinusoidal shape. \cite{Zharikov2013} searched the period of the V/R  using a Discrete Fourier Transform method, and the dominant frequency they found in the power spectrum was 0.01187 cycles/days, corresponding to a period of 84.2 days, which agrees with our results. Our result in this work agrees with the period of the orbital system of 84.1 days found by \cite{Bjorkman2002}. 

\subsubsection{Double-peak separation variability}
By measuring the DPS, information about the rotational velocity, size of the emitting region, inclination, temperature and density distribution, and disk dynamics can be inferred. Because DPS is primarily a result of Doppler shifts due to the Keplerian rotation of the disk, faster rotation leads to a larger separation, implying that the emission originates closer to the star in the inner disk \citep{Huang1972}. We note that this is also sensitive to the inclination angle of the disk relative to the observer, whereas, for edge-on disks, the separation is wider due to the higher radial velocity components visible to the observer. For $\pi$~Aqr, the DPS curve goes from a high value, close to 530~km/s, to a lower value of $\sim$ 100~km/s, in all analyzed periods.
In detail, from December 2001, the DPS continuously drops its value until October 2011 (second gray vertical line), where it remains almost constant at around 300~km/s. After the inflection point, the DPS decreases continuously until it reaches an almost constant value around $\sim$ 150~km/s. Therefore, based on H$\alpha$ measurements, the emitting region increased $\sim$ 18 times its value since the inflection point and decreased by nearly twice its rotational velocity. 

\subsection{Other lines}
As we mentioned in the observational data section, not all observations downloaded from BeSS and BeSOS contain the full optical range. Therefore, it is not possible to compare the other studied lines from Table~\ref{tab:lines} with the exact five selected epochs of H$\alpha$. Hence, from the 16 spectra covering the optical region, we chose 14 dates to compare simultaneously the shape of the lines (the remaining two were not chosen because they were at consecutive dates where no considerable variation was seen). The selected elements presented emission at least once: helium, silicon, and iron, as well as H$\beta$ and H$\gamma$. From H$\alpha$ line, we infer that all chosen dates are in a disk phase. The line profiles are presented in velocity, and the intensity was normalized to the continuum. We remember that the lines are all corrected by the contribution of the atmosphere. The related figures are displayed in the Appendix \ref{app:evolbalmer} for Balmer, Appendix \ref{app:evolhelium} for helium, Appendix~\ref{app:evolsi} for silicon, and Appendix~\ref{app:evoliron1} for iron. In the following, we compare the shape of the line profiles in detail concerning the Balmer lines. We note that three of the five disk phases discussed above coincide with the observation dates chosen here for the other lines (October 09, 2018, and November 25, 2022). 

\subsubsection{H$\beta$ and H$\gamma$ hydrogen lines}
Figure~\ref{app:evolbalmer} shows H$\alpha$, H$\beta$, and H$\gamma$ line profiles for different disk phases. The vertical axis is different for H$\alpha$ due to its strong emission. From this figure, we see that when the emission is presented, a DPS is always observed; however, the V/R intensity ratio is not necessarily equal for all Balmer lines, and the intensity of the emission is not always descending with respect to higher orders in the series (higher-order Balmer lines require higher densities and temperatures to emit significantly; thus they primarily trace the inner, denser disk). For example, in October 2011 (which also coincides with one analyzed disk phase at the second gray vertical line), H$\beta$ is less intense than H$\gamma$. This repeats in November 2020 and September 2022. Moreover, on these three occasions, the DPS is also different, and it seems that H$\beta$ does not follow the same symmetry as H$\alpha$ and H$\gamma$. This suggests that the regions of the disk contributing to the emission of these lines differ significantly in their optical depth and/or geometry. Following the H$\alpha$ emission line shape and the results for the V/R intensity ratio, a small overdensity is presented in the disk for years until the inflection point occurs (October 2014), where the disk density increases and its distribution changes.

\subsubsection{Helium lines}
Figure~\ref{app:evolhelium} shows three selected helium lines from the Table~\ref{tab:lines}, which show emission lines features. These helium lines are HeI $\lambda$ 5875, 6678, and 7065 \AA. An empty box will appear if no data are available for the line on that date. In general emission lines show a shell profile with low intensity emission ($\leq$ 1.1). We note after the inflection point, between 2016 and 2020, no emission (or a very low) is seen in $\lambda$ 6678 and 7065 \AA\,, but a persistent low emission is observed in 5875 \AA. This line is more easily excited than the other two HeI lines, which are higher-energy transitions. Then, if the disk is dense and moderately heated but not extreme (we obtained a $\rm T_{eff} \sim $24000 K, and if we consider that $\rm T_{disk}$ is between 60 and 72\% of $\rm T_{eff}$ then  $\rm T_{disk}$ is approximately between 14000 and 17000 K) the 5875 \AA\ line, may appear in emission while the others do not. Denser and higher temperature disk regions are found close to the star; therefore, HeI emission lines are traced to the inner part of the disk or optically thick disk zones. The last observation date in Figure \ref{app:evolhelium}, when H$\alpha$ is higher in intensity and larger in EW, shows emission in the three HeI lines, with similar intensities and with a V/R < 1, as observed in Balmer lines.

\subsubsection{Silicon and iron lines}
Figure~\ref{app:evolsi} shows the silicon lines 6371 and 6347 \AA\, and Fig.~\ref{app:evoliron1} displays the iron lines 4232, 4584, 5018, 5169, 5198, 5235, 5276, 5316, 5363 and 6385 \AA. The calculated values of the EW and DPS for spectral lines of both elements are available in the \href{https://docs.google.com/spreadsheets/d/1jDAFGvOOEU8OPzYMJQ7btGLXoisBngnSnmS6OdfEo8A/edit#gid=764686422}{online excel file}. 
Although the spectra of SiII lines are very noisy, a very low emission is distinguished between 2019 and 2022. The emission lines are asymmetric, showing a DPS. These low-ionisation lines form in cooler and denser regions of the disk compared to helium or hydrogen lines.  

All FeII lines have very similar behavior except for the case of FeII~5018 \AA \,, where an absorption line can be distinguished in the spectra between 2001 and 2013, with a small DPS profile in some observation epochs, similar to H$\beta$. Since emission appears in FeII~5316 \AA, it is the most prominent compared to the other iron lines.
Although the V/R is more similar to H$\beta$ than helium lines, the shape of the lines shows a flat structure in the emission peak, with a third peak. In 2022, FeII $\lambda$ 5018 and 6385 \AA\, show opposite behavior in the V/R ratios compared to the other iron lines. Not all the FeII data exhibited V/R  variations over the years. In some lines, no emission or an absorption scheme is observed. The most significant variation of the DPS is encountered for FeII~5018 \AA, where the highest DPS is 585 km/s on 2010-10-26. Compared with H$\alpha$ on the same day, it is almost twice as large. Following this, the DPS continuously drops its value to its minimum of 152~km/s. From FeII~5169 to 5363 \AA, the DPS fluctuates from 412 to 205~km/s. The DPS of FeII~5198 \AA \, continuously decreases from 372 to 204~km/s. All the other lines have variations between their maximum and minimum peaks. We remark that the highest value of DPS for each line occurred on 2015-10-31, with the highest DPS of 412~km/s for FeII~5276 \AA.  From FeII~5169 to 5276 \AA \, the smallest DPS is observed in 2022-09-03. Any FeII line has a similar behavior decrease in the DPS compared to H$\alpha$, except for the FeII~5198 \AA. The lowest DPS is found for FeII~6385 \AA, equal to 87 km/s. This is the smallest DPS for all the lines in $\pi$ Aquarii. For FeII~6369 and 6385 \AA, the DPS is decaying similarly to H$\alpha$. In the discussion section below, we include Fig.~\ref{fig:dpsewfeii} with the EW versus DPS of iron emission lines studied in this work. 

Since the inflection point, almost all iron lines have been in emission. FeII emission is generally associated with intermediate-density regions of the disk, typically closer to the outer or mid-disk. Its transitions involve lower excitation energies and are less temperature dependent than SiII. By comparing the variations of the spectral lines of the other elements, the strong hydrogen and FeII emission lines, alongside weak SiII emission, suggest that the outer disk is dominant (the outer and cooler parts of the disk are contributing most of the observed line emission). 

\section{Discussion}
\subsection{Possible effects of not considering continuum contribution in the equivalent width determination}
As we mentioned previously, most of the spectra used in this work are already normalized to the continuum in the BeSS and BeSOS databases. This is relevant because during an episode of mass ejection, the disk can grow, and the distribution of material in the disk can change. %The additional material may increase the density of the disk, leading to greater scattering of the star's light and potentially more absorption or emission in the spectrum. 
On the one hand, when the Be star is viewed edge-on, a thicker disk with more ejected material would cause a greater reduction in the continuum intensity because the material in the disk can block or scatter more of the star’s light. The continuum intensity may decrease, and the spectral features (e.g., emission or absorption lines) may become more prominent. On the other hand, when the system is viewed pole-on, the observer sees less of the disk material, and the scattering effects are reduced. The continuum observed would be closer to the star’s intrinsic continuum, and changes in the observed continuum due to mass ejection into the disk would be less pronounced. Emission lines may also be less affected because the amount of disk material in the observer's direct line of sight is smaller.
Therefore, the EW could be underestimated by the changes in the continuum, overall, if the Be star is seen from the edge-on. We note that we discuss and conclude relevant aspects based on the EW values obtained without taking the continuum into consideration. In particular, when we claim that the disk is dissipating (because the EW is lower), it may not be completely true since the effect described above. 

\subsection{Relation between helium, silicon, and iron lines concerning the disk phases}
\begin{figure}
\begin{center}
\includegraphics[width=\columnwidth]{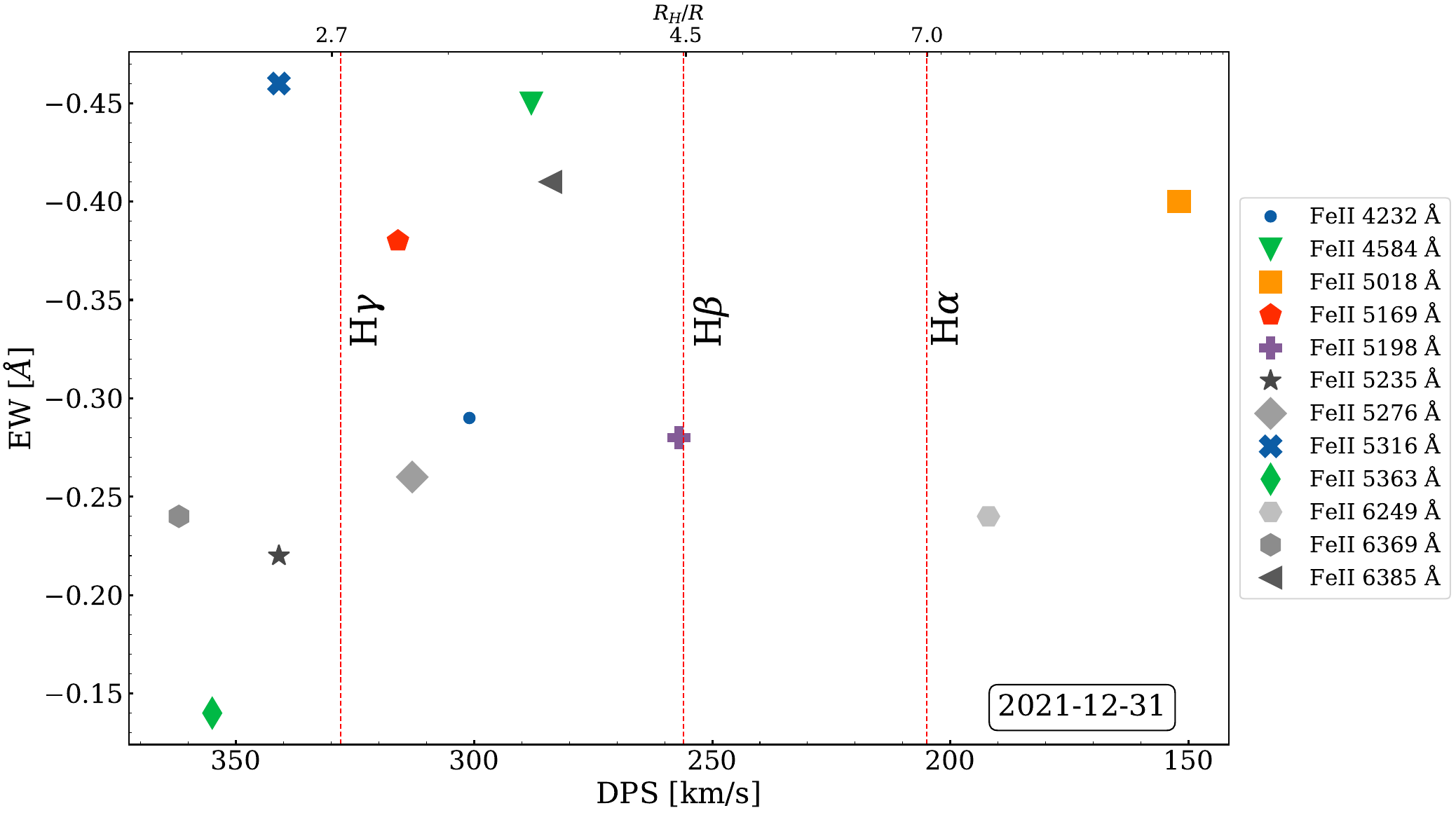}
\caption{Double-peak separation of iron lines compared to their EW for the penultimate observation date of Fig.~\ref{app:evoliron1}. This date corresponds to} December 31, 2021. Red vertical dashed lines denote the DPS for the Balmer lines at the same date. 
\label{fig:dpsewfeii}
\end{center}
\end{figure}

Many iron lines present emission since the inflection point of H$\alpha$. Several works in the literature are related to the study of iron lines (e.g., \cite{Hanuschik1987,Ballereau1995,Arias2006}). The first work divides the shape of the line profile of iron lines into two classes: Class 1 are double-peaked and symmetrical emission lines, which describe the relationship between the width of the peaks concerning $v\sin i$. Class 2 line profiles exhibit single asymmetric sharp peaks in a complex, broad emission structure. It also concludes that the emitting region of iron lines is inside H$\alpha$ and H$\beta$ formation region zone, close to the central star and that they are optically thin. \\
It is possible to calculate the emitting region for spectral lines using the \cite{Huang1972} relation:

\begin{equation}\label{eq:huang}
   \rm  R_H=\left ( \frac{2\,v\sin i}{DPS} \right )^{2}R,
\end{equation}

\noindent where $v\sin i$ is the equatorial projected rotational velocity of the central star, $\rm R$ is the stellar radius, and $\rm R_H$ is Huang's radius
for the emitting region. This relation was used to compute Huang's radius for the Balmer lines on the penultimate observation date of Fig.~\ref{app:evoliron1},
which correspond to 2021-12-31, where $\rm v\sin i = 271\, km/s$, taken from Table~\ref{tab:HD212571}. We have for H$\alpha$: $\rm R_H=7.0 R$, %\textbf{(which conforms to the emitting region between 4 and 12 stellar radii described in Figure 7 in Rivinius et al. \citeyear{Rivinius2013})},
H$\beta$: $\rm R_H=4.5 R$ and H$\gamma$: $\rm R_H=2.7 R$. Looking at the FeII lines, the only ones outside the emitting region of H$\alpha$ and H$\beta$ are FeII~5018 and 6249 \AA, with $\rm R_H$ equal to 12.7 and 7.9 $\rm R$, respectively. We also note that FeII~5018 \AA \,is the only FeII emission line showing a different shape profile (V/R) compared to the rest of the lines (see Fig.~\ref{app:evoliron1}). In Fig.~\ref{fig:dpsewfeii}, we show the variation between EW and DPS for 2021-12-31 of iron emission lines indicated with different colors and symbols. The plot also shows the DPS of Balmer lines to compare with. We chose this date because all FeII emission lines show a clear DPS and have spectra on that observation epoch. Looking at the figure, we can conclude that there is no correlation between the EWs of iron lines and the disk formation region. The DPS shows a wide variation range. Almost all iron lines are concentrated on the inner disk inside the H$\beta$ emitting region.\\

\cite{Ballereau1995} asserted that iron lines must be optically thick, which can change the location of the formation region where FeII lines are formed. \cite{Arias2006} studied the optical depth regime of FeII $\lambda$ 5198, 5235, 5265, 5276, 5317, and 5365 \AA \, lines and the degree of expansion of their formation region simultaneously, finding out that these lines are optically thick. They are formed in circumstellar regions close to the central star. This is also confirmed by the fact that the source function of FeII lines rapidly decreases with radii, preventing the formation in the outer part of the disk. They calculated the extent of the emission region of FeII lines using the self-absorption-curve (SAC) method \citep{Friedjung1987}, which allows the opacity eﬀect to be explicit on the emitted radiation intensity. This is in contrast to Huang relations, which are only for optically thin lines and result in uncertainty in the outcome presented before.
\subsection{Evolutionary status of $\pi$\,Aquarii}
\begin{figure}
\begin{center}
\includegraphics[width=\columnwidth]
{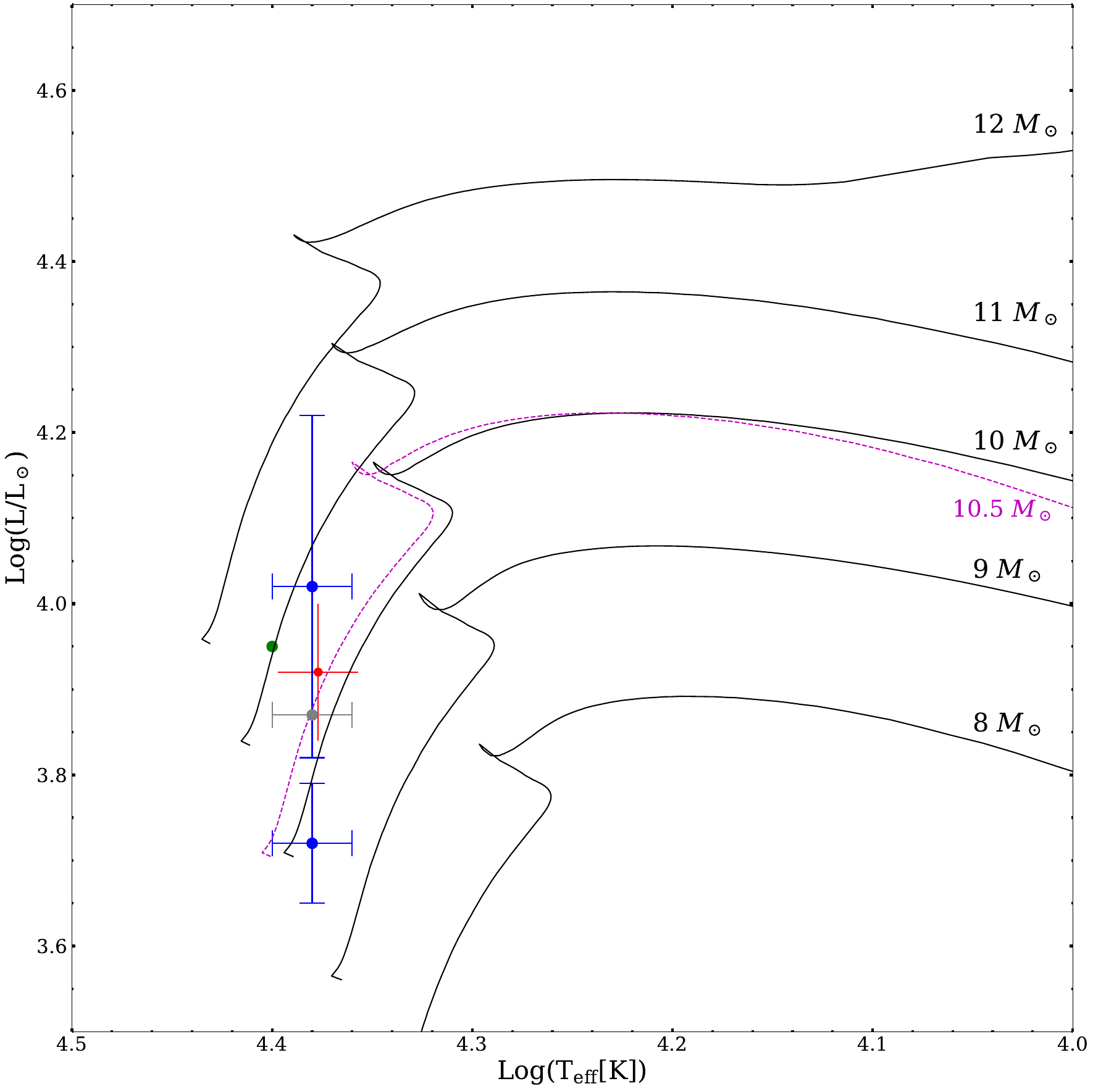}
\caption{Hertzsprung-Russell diagram for $\pi$\,Aquarii. The solid lines represent the evolutionary tracks of rotating stars with solar metallicity and $\omega=0.8$, as outlined by \citet{Georgy2013} and \citet{Granada2013}, the dash magenta line has an initial $\rm M=10.5\ M_\odot$. The filled red circle denotes the findings from this study. The filled blue circle indicates the luminosity and temperature derived from \citet{Zharikov2013}, calculated using varying distances. The filled gray circle corresponds to data from \citet{Tsujimoto2023}. The filled green circle corresponds to data from \citet{Hohle2010}.}
\label{fig:georgymodels}
\end{center}
\end{figure}

We study the evolutionary state of $\pi$ Aquarii with the results obtained with the code \texttt{ZPEKTR}, using the evolutionary tracks from \cite{Georgy2013,Granada2013}. This grid was made for a range of masses between 0.75 to 15 M$_\odot$, separated in 270 evolutionary tracks, for rotating stars at $\rm Z = 0.014,\, 0.006\, and\, 0.002$, the interpolated models were obtained from
\href{https://www.unige.ch/sciences/astro/evolution/fr/base-de-donnees/syclist//index/}{this link}.
Figure~\ref{fig:georgymodels} shows the Hertzsprung-Russell diagram for stars with $\rm \omega =0.80$ and $\rm Z=0.014$, where the solid lines indicate the evolution tracks with the corresponding initial mass. \cite{Zharikov2013} computed the mass using the stellar parameters; they calculated the M$_V$ with system brightness V equals 4.85 mag, $\rm A_V=0.15$ mag, and HIPPARCOS distance of 340 pc, and using a $\rm T_{eff} =24000$ K with a bolometric correction $\rm BC = -2.36$ mag, they finally obtained  $\rm \log{L} = 4.02\ L_\odot$ that leads to an initial mass of 10.5 M$_\odot$. Using a shorter distance of 240 pc \citep{Leeuwen2007}, the initial mass changes to 9.5 M$_\odot$ with the models from \cite{Ekstrom2012}.

Our result is consistent with an initial mass between 10 and 11 M$_\odot$, located within the main sequence. This is about halfway through the main sequence. This result agrees with the result of \texttt{ZPEKTER}. The dynamical mass found by \cite{Bjorkman2002} is between 12.5 and 17 M$_\odot$, calculated in a disk-less phase with a range of $i$ from 65$^\circ$ - 85$^\circ$. Our stellar parameters were computed in 2013, when the EW of H$\alpha$ dropped to a minimum, with a range of $i$ between 50$^\circ$ - 75$^\circ$. With these new models, the initial mass of \cite{Zharikov2013} would change from 10 to 11 M$_\odot$. \cite{Tsujimoto2023} using $\rm T_{eff} = 24000 \pm 10000$ and a Gaia distance of 286 pc obtained a value of $\log \rm L = 3.87\ L_\odot$, placing it in the evolutionary state with an initial mass close to $\rm 10.5 M_\odot$, where the range they proposed between 9 and 15 $\rm M_\odot$ is consistent, the reason for this wide range is to include the dynamical and evolutionary mass.

Several authors have calculated $\rm \log g$ and $\rm v\sin i$ of $\pi$ Aquarii. \cite{Chauville2001} studying the HeI~4471 \AA \, line determined $\rm T_{eff}$ and $\rm \log g$ of 116 Be stars. For $\pi$ Aquarii they found $\rm T_{eff} = 26668$ K and $\rm log g =3.95$ dex. \cite{Fremat2005}, utilizing a code that considers the fast rotation of stars as rigid rotators and gravity darkening, found  $\rm T_{eff} = 26061\pm736$ K and $\rm \log{g}= 3.92\pm0.09$ dex. \cite{Arcos2018} used photometry data fitted with Kurucz \citep{Kurucz1994} and TLUSTY \citep{Hubeny1995} stellar atmosphere models, together with spectroscopy data, to find the parameters of a Be star sample. For our star the results were: $\rm T_{eff} = 24500\pm245$ K, $\rm \log g = 3.4\pm0.03$ dex and $\rm v\sin{i} = 215\pm4$ km/s. \cite{Ahmed2017} using photospheric nitrogen abundances, computed $\rm T_{eff} = 26061$ K and $\rm \log g = 3.7$. We note that different authors' temperatures are similar, but the effective gravities in all works are always lower than 4.

\cite{Rivinius2013} summarize the main properties of the rotation of Be stars. They derive an average value of $\rm \bar{W}=0.8$, based on spectroscopic \citep{Fremat2005} and interferometric \citep{Meilland2012} analyses. They note that this value does not depend on temperature or effective gravity; more importantly, the lowest $W$ value for a B star to become a Be star is about 0.7. The $\rm W$ calculated in our work is lower than these values, about 22$\%$. Looking at other works, \cite{Zorec2016} computed the true rotational velocities of 233 Be stars, fixed by the effects of gravitational dimming, binarity, and macroturbulence, and found that the mode of $\gamma$ is equal to 0.65. The value of $\gamma$ of $\pi$~Aquarii in Table~\ref{tab:HD212571} agrees with this value. \cite{Cochetti2019} derived the mean values of $\gamma=0.75$ and $\omega=0.9$ of 18 Be stars using interferometric observations. In comparison with the latter authors, it is found that the values of $\gamma=0.63$ and $\omega=0.82$ calculated by \texttt{ZPEKTR} of $\pi$~Aquarii are smaller. 

\begin{figure}
\begin{center}
\includegraphics[width=\columnwidth]{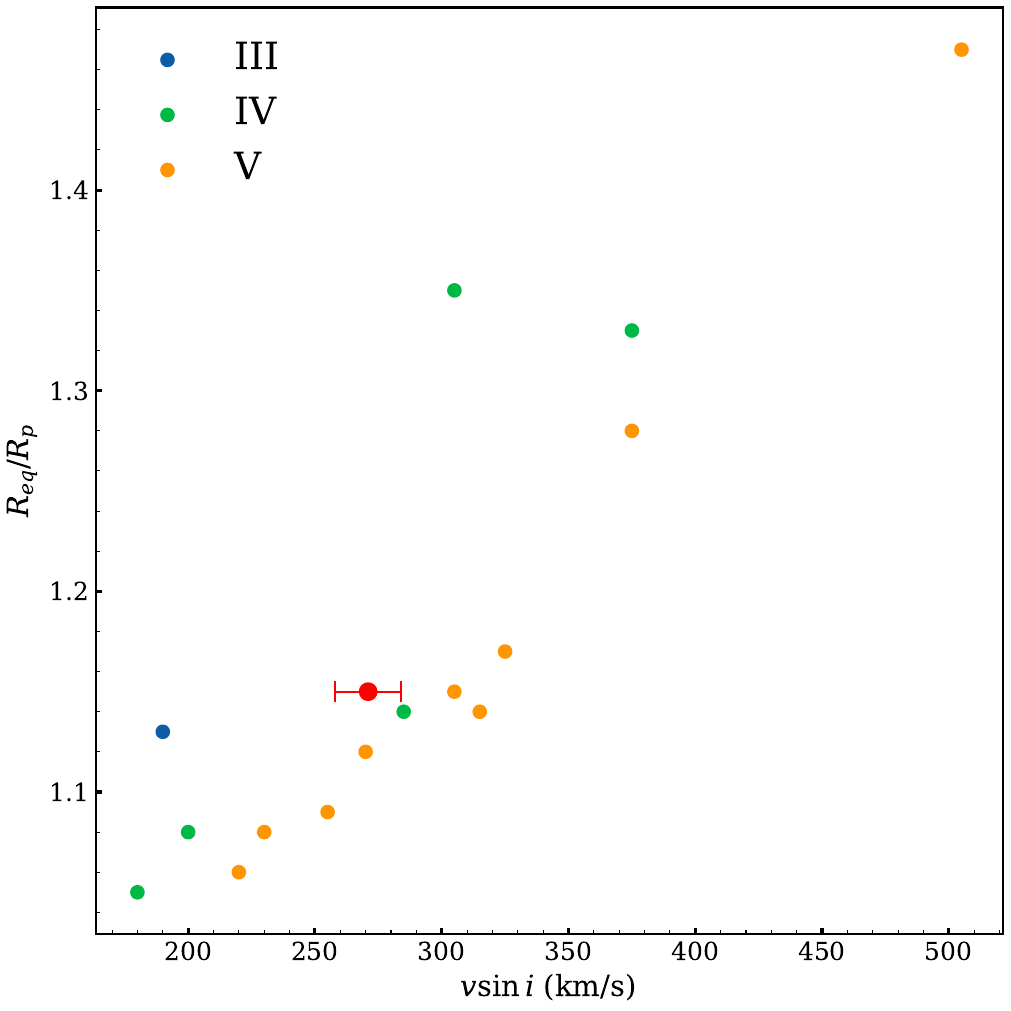}
\caption{Oblateness versus projected velocity of Be stars, as presented by \citet{2012Vanbelle}. The solid circle represents $\pi$ Aquarii, with data derived from Table~\ref{tab:HD212571}. Different colors indicate the luminosity class of each star, with the red color representing the star Pi Aquarii.}
\label{fig:vonbelle}
\end{center}
\end{figure}

Figure~\ref{fig:vonbelle} shows the $\rm R_{eq}/R_p$ ratio versus projected velocity for different Be stars from \citet{2012Vanbelle}, ranging from III to IV in luminosity class, which is in agreement with the spectral classification of $\pi$ Aquarii, which is classified as B1III-IVe by \citet{1982Slettebak}.

\subsection{Quantification of the disk changes under parametric model assumptions}
We noticed that the profiles of this star range from shell type to DPS to triple-peaked to flat-topped profiles. This is typical behavior for a Be star in a binary system \citep{Okazaki2002,Marr2021,Panoglou2016,Panoglou2018,Cyr2017}. \cite{Panoglou2018} used a smoothed particle hydrodynamics (SPH) code to compute the configuration of Be discs in coplanar circular binary systems. The output from SPH consisted of a disk configuration of two spiral arms. This result was used as input into \texttt{HDUST} code to obtain the theoretical H$\alpha$ emission lines view at different inclination angles. They proposed that triple-peaked profiles are an evolved state of flat-topped profiles. Moreover, they stated that the evolution with the orbital phase of the profile shape can be illustrated as a DPS followed by a flat-topped profile then a triple-peaked profile and again a flat-topped profile and then a DPS profile. Then, $\pi$ Aquarii can be in such a coplanar or misaligned binary system \citep{Moritani2013}. 

By assuming the VDD's parametric model, we obtained good results in modeling the H$\alpha$ emission line for the epochs 2001, 2011, and 2014. This last
presented a bimodal distribution in all parameters except for the inclination angle. In 2001, the superficial density was lower and decayed faster, close to the
star. In 2011 and 2014, the density decays slower with $\rm n$ exponent values n $\sim$ 3.2. In the last two epochs, 2018 and 2022, the model fit the intensity
of the profiles but not the shape. On one hand, in 2018 the emission line is irregular with a flat-topped peak, on the other hand, in 2022 the emission line
presents a DPS profile with R$\gg$V. Also, the wings are not well reproduced, mainly because \texttt{HDUST} does not reproduce electron scattering (a process
that widens the wings of the line). For these two epochs, we can see that the n exponent does not converge correctly, giving the lowest possible value of n
within the grid (n $=$ 3.0). Smaller values of n are needed to achieve a slower density decay near the disk to reproduce the observed line. The superficial
density increases with the years, as can be observed visually as a growth in the shape of the emission line. The inclination angle ranges between $\sim$ 60$^{\circ}$ and 77$^{\circ}$ (or 70$^{\circ}$ if we consider only the acceptable fits). The possible reason for such a difference in the change of the inclination angle is that the disk is precessing and therefore is in a misaligned binary system.\\

\citet{Cochetti2019} used HDUST to model infrared spectra obtained in June 2017 with FIRE spectrograph and obtained good fits with the following parameters. $\rm n$ = 3.5, $\rho_{0}$ = 10$^{-11}$ $\rm g\,cm^{-3}$ and $\rm i$ = 45$^{\circ}$. Comparing these results with the parameters obtained by us at a similar epoch (2018-10-09) (see Table~\ref{tab:Disk_emcee}), we found no similarities for the angle of inclination and the value of $\rm n$.
These differences may be related to inaccuracies in the modeling of the H$\alpha$ line, as mentioned in the previous paragraph, and also by the approximations for the mass, radius of the star, and rotation rate used to create the infrared models. As for the parameter $\rho_{0}$, calculated using Eq. 5, we found a value of 1.75\,X\,$10^{-11}$ $\rm g\,cm^{-3}$, slightly higher than that estimated by \citet{Cochetti2019}

\subsection{Origin of X-ray emission in $\pi$ Aquarii: Accretion onto the secondary or interaction in the Be disk}
\begin{figure}
\begin{center}
\includegraphics[width=\columnwidth]{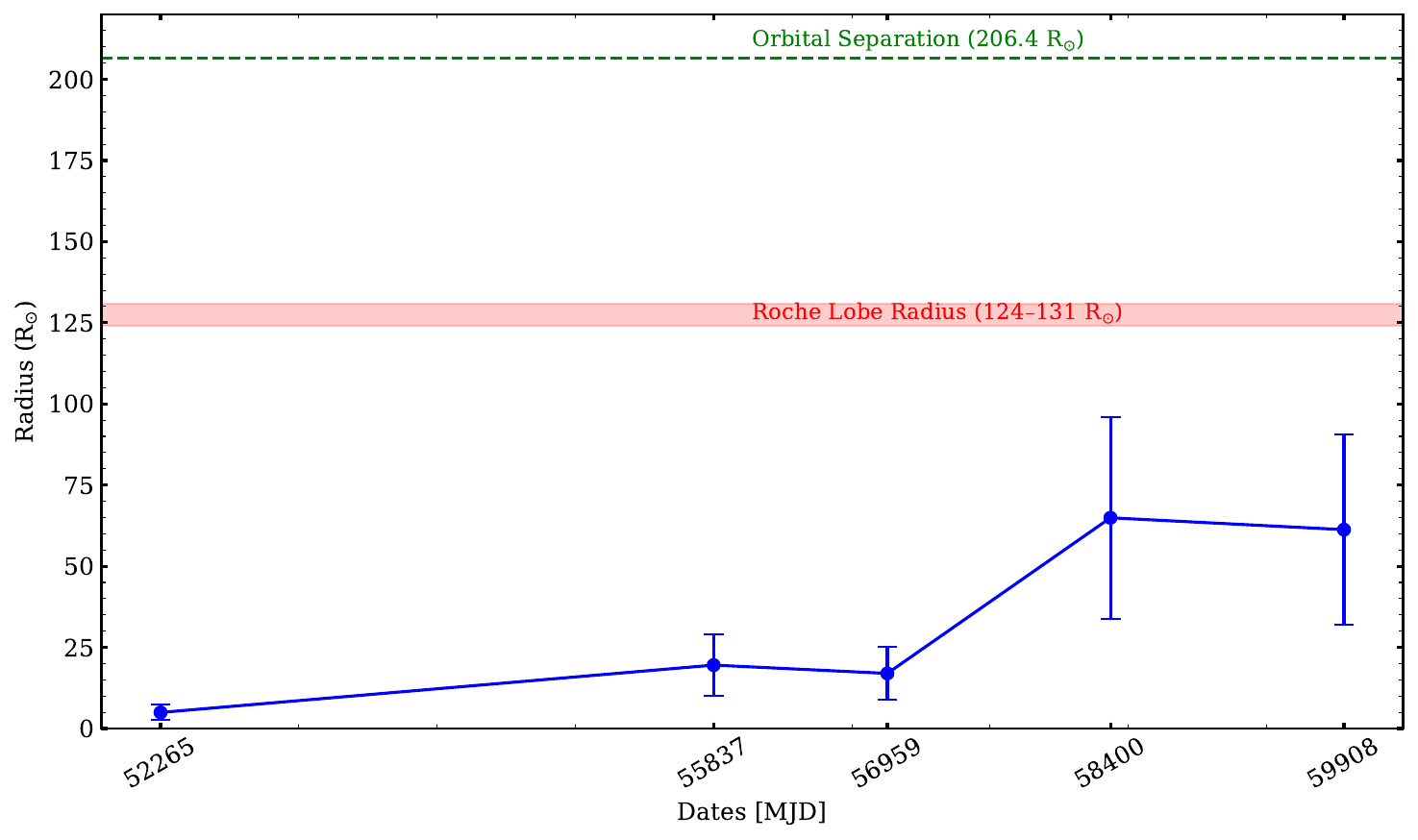}
\caption{Evolution of the disk radius of $\pi$ Aquarii derived from H$\alpha$ emission line measurements using Huang's law. The disk radius (blue points with error bars) shows significant growth across five epochs, increasing from 4.9 to 64.9 R$_{\odot}$. The Roche lobe radius (red shaded region, 124--131 R$_{\odot}$) and orbital separation (green dashed line) indicate that the disk remains well within the Roche lobe and is not truncated by the companion star. Errors represent uncertainties in the Huang radius.}
\label{fig:expndisk}
\end{center}
\end{figure}

As mentioned in the introduction section, $\pi$ Aquarii presents hard continuous thermal X-ray emission, which can be the product of accretion onto the secondary (most probably a magnetic or non-magnetic WD),  wind or magnetic interaction between the star and its decretion disk \citep{Smith2016,Rauw2022}.  
Our five selected dates mark a change in the EW H$\alpha$ variation; therefore, we interpret these changes as a disk phase. We can obtain an estimation of the emitting region by using Huang's law (see eq.~\ref{eq:huang}). From our measurements, the DPS of the H$\alpha$ emission line varies significantly across the five epochs, ranging from 563.7 km/s to 155.3 km/s. By fixing the $v\sin{i}$ value to 271.13 km/s and using R = 5.33 R$_{\odot}$ (see Table~\ref{tab:HD212571}), we estimate that the emitting region of the disk expands from 4.9 to 65.0 R$_{\odot}$, indicating a significant growth of the disk over time. This growth is likely driven by an enhanced mass loss from the primary star. We computed the Roche lobe radii for the binary system using the formulation of \cite{1983Eggleton} with  M$_1$= 11 M$_{\odot}$ (see Table~\ref{tab:HD212571}) and M$_2$ between 0.5 and 0.8 $M_{\odot}$ \citep{Tsujimoto2023,2024Klement,Huenemoerder2024} and found that the disk remains well within the Roche radius (\( R_L \approx 124\text{--}131 \, R_{\odot} \)). This implies that the companion star should not have accreted significant mass from the disk, at least by overflow mass transfer mechanism, as the disk has not reached the Roche lobe boundary (see Fig. \ref{fig:expndisk}). Furthermore, considering the orbital separation of 0.96 AU (206.4 R$_{\odot}$ proposed by \cite{Bjorkman2002}, we conclude that the companion star does not truncate the disk (emitting region of the optical range) of the Be star.

%The lack of disk truncation by the Roche radius may also explain the lack of correlation between X-ray emission and H$\alpha$ variability, as the accretion rate onto the companion star could remain relatively stable despite changes in the disk structure.
If there is no accretion by overflow mass transfer mechanism, the X-ray emission could come from the interaction between the star and its decretion disk or accretion by other mechanisms. \cite{Naze2019a} compared the EW H$\alpha$ and X-ray fluxes from simultaneous data taken from April to December 2018, finding no correlation over time (unusual for a $\gamma$ Cas star). Later, \cite{Huenemoerder2024} obtained six high-resolution X-ray spectra of $\pi$ Aquarii with the Chandra/HETG spectrometer from August to October 2022. Their analysis indicates the presence of very hot thermal plasma and coherent variability in the light curve. In five out of six observations, they found a nearly stable hardness ratio ($1.50\times10^{-11}$ erg/cm$^{2}$s on August 23 and 27; September 05, 13 and 14), whereas the final observation (October 30) showed a significantly harder ($2.54\times10^{-11}$ erg/cm$^{2}$s) and more absorbed spectrum. They attribute the X-ray emission to accretion onto a (magnetic) WD. This means that the magnetic field of the WD could channel material directly from the wind of the Be star or from the edge of its disk without the Roche lobe needing to be full. Moreover, as we concluded before, the disk of $\pi$ Aquarii is precessing. This means that there are two main scenarios for the accretion to occur: (i) the WD crosses the disk at two points in its orbit, and it is at these moments that it captures material and temporarily increases its X-ray emission, or (ii) if the orbit is very steep, the WD spends most of its time outside the disk. In this case, it can only accrete material from the wind of the Be star, and the accretion rate is much lower than if it had passed through the disk. Following our results, the inclination angle varied $\sim$ 10$^{\circ}$ in 20 years, indicating that the first scenario is more probable, and we should expect small increases in X-rays as the WD crosses the disk. On the other hand, as we noted before, FeII~5018 \AA \, covers an emission region larger than H$\alpha$, and is the only FeII emission line showing a different shape profile (V/R) compared to the rest of the lines. This could be due to changes in the disk geometry and its alignment with the WD.

\section{Conclusions}
A rapid rotator Be star in a binary system, $\pi$ Aquarii has been studied since the 1930s due to its conspicuous variability. This work aimed to reveal new information on the spectral variability of this star as well as the stellar and disk parameters by characterizing its behavior between several spectral lines that have changed between absorption and emission features over the years. We used observations from such star catalogs as BeSOS and BeSS and obtained the EW and V/R for Balmer, iron, silicon, helium, and other lines. All data are available in an \href{https://docs.google.com/spreadsheets/d/1jDAFGvOOEU8OPzYMJQ7btGLXoisBngnSnmS6OdfEo8A/edit#gid=764686422}{online Excel file}. \\

Our physical parameter estimations based on the spectrum fittings allowed us to obtain good agreement 
with other works on $\pi$~Aquarii. Our mass estimates agree with the lower limit of dynamical mass determined by \cite{Bjorkman2002}. Moreover, the estimates on temperatures, gravities, and $v\sin{i}$ are compatible with the literature. In this work, we still assumed oblateness of the star due to its fast rotation, obtaining values for polar and equatorial stellar parameters. \\

Regarding the disk parameters, we studied five epochs that show a large change in the EW in the years 2001, 2011, 2014, 2018, and 2022. We found that the density of the disk decays faster (with n $\sim$ 3.9) in 2001. In this epoch, the disk was not prominent, showing a shell profile in the H$\alpha$ emission line. Then, in 2011 and 2014, the disk was growing, with a density distribution decaying more slowly (n $\sim$ 3.2). In the last two epochs, 2018 and 2022, the H$\alpha$ emission line presents a very broad shape with high intensity (almost double the other epochs), and we could not obtain good agreement, mainly because or method does not take into account the electron scattering in the disk and also because we need lower values (n $<$ 3.0) for the n exponent decay of the density distribution (to obtain a much slower decay) to improve our H$\alpha$ fitting. We assert that the disk of $\pi$ Aquarii is in a misaligned binary system, and its inclination angles change between $\sim$ 60$^{\circ}$ and 70$^{\circ}$, going through shell profiles to DPS to triple-peaked to flat-topped profiles, and it now shows a DPS profile (with R$\gg$V). We expected to implement the SAC method utilized in \cite{Arias2006} to compute the emitting region of the emission lines and compare them with our results using Huang's relation.

Following our findings, we propose that the (magnetic) WD crosses the disk at two points in its orbit, and it is at these moments it captures material and temporarily increases the X-ray emission. Also, we found that FeII~5018 \AA \, covers an emission region larger than H$\alpha$ and that it is the only FeII emission line showing a different shape profile (V/R) compared to the rest of the lines, indicating changes in the outer disk probably related to the WD. 

The star $\pi$ Aquarii remains an unsolved case of $\gamma$ Cas type stars, and further studies are needed to fully understand what is happening with this system. Simultaneous UV and X-ray observations could reveal correlations between changes in the stellar wind and X-ray emissions.

\section*{Data availability}
The data used in this work are available on Zenodo at \url{https://doi.org/10.5281/zenodo.15039731}

\begin{acknowledgements}
DC is grateful for the financial support from Fondecyt N.1231637. CA, MC, and IA are grateful for the financial support from Fondecyt N.1230131. TS is grateful for the financial support from Fondecyt N.3230770. MC \& CA acknowledge the support from Centro de Astrof\'isica de Valpara\'iso. DT acknowledges support from \emph{ANID BECAS/DOCTORADO NACIONAL/2024-21240465}. This research was partially supported by the supercomputing infrastructure of the NLHPC (ECM-02). This project has received funding from the European Union’s Framework Programme for Research and Innovation Horizon 2020 (2014-2020) under the Marie Skłodowska-Curie Grant Agreement No.~823734 and is also Co-funded by the European Union (Project 101183150 - OCEANS). This work used BeSOS Catalog, operated by the Instituto de F\'isica y Astronom\'ia, Universidad de Valparaíso, Chile: http://besos.ifa.uv.cl and funded by Fondecyt iniciación 11130702. This work has used the BeSS database, operated at LESIA, Observatoire de Meudon, France: http://basebe.obspm.fr. This work has been possible thanks to the use of AWS-U.Chile-NLHPC credits. Powered@NLHPC: This research was partially supported by the supercomputing infrastructure of the NLHPC (ECM-02). 
\end{acknowledgements}

%\section*{Data Availability}
%The article's data is available in an online format and observation data can be shared on request with the corresponding author.

%\begin{references}
%\begin{thebibliography}{}
\bibliographystyle{aa} % style aa.bst
\bibliography{references} % your references Yourfile.bib
%\end{references}
%\end{thebibliography}

\begin{appendix}
%%%%%%%%%%%%%%%
\onecolumn
\section{Instrument and resolution of BeSS database}

\begin{longtable}{lllll}
\caption{Instruments of the spectra used from BeSS for the optical range and H$\alpha$ (individual observations).} 
\label{tab:bess_optical_individual} \\
\hline\hline
Spectrograph & Date & Resolution & Range [\AA] & Orders \\
\hline
\endfirsthead
\caption[]{(continued)} \\
\hline
Spectrograph & Date & Resolution & Range [\AA] & Orders \\
\hline
\endhead
\hline
\multicolumn{5}{r}{\textit{Continued on next page}} \\
\endfoot
\hline
\endlastfoot

% Data from first table
\noalign{\smallskip}
% ELODIE & 2001-12-22 & 45000 & 3900-6810 & 67 \\
eShel & 2008-07-21 & 10000 & 4380-7000 & 19 \\
eshel101 & 2009-08-19 & 10000 & 4360-6700 & 18 \\
eShel & 2010-10-26 & 10000 & 4300-7100 & 21 \\
eShel\_T\_Garrel & 2011-10-03 & 11000 & 4300-7300 & 22 \\
eShel\_T\_Garrel & 2012-09-22 & 11000 & 4140-8100 & 27 \\
Eshel & 2013-09-20 & 11000 & 4200-7300 & 23 \\
eshel-5 & 2014-10-29 & 11000 & 4280-7100 & 21 \\
Eshel & 2015-10-31 & 11000 & 4200-7200 & 23 \\
Eshel & 2016-10-31 & 11000 & 4200-7200 & 23 \\
Eshel & 2017-10-31 & 11000 & 4200-7200 & 23 \\
eShel101 & 2018-10-09 & 11000 & 4200-7200 & 23 \\
eShel101 & 2019-10-16 & 11000 & 4200-7200 & 23 \\
Eshel + ESP & 2020-10-30 & 30000 & 3750-7550 & 31 \\
NOU\_T & 2021-12-31 & 8500 & 3700-8100 & 33 \\
NOU\_T2 & 2022-09-03 & 8500 & 3700-8800 & 35 \\
\hline\hline
\noalign{\smallskip}
Spectrograph & Number of Spectra & Resolution & \multicolumn{2}{c}{$\rm H\alpha$} \\
\hline
\noalign{\smallskip}% Data from second table
Eshel & 29 & 10000, 11000 & -- & -- \\
F/10 & 1 & 6570 & -- & -- \\
L-200 & 7 & 5950, 5865, 5881, 5962, 5811, 5755 & -- & -- \\
L200 & 2 & 9864, 8033 & -- & -- \\
L200-1800 & 1 & 11208 & -- & -- \\
L200\_1800 & 11 & 7819, 7852 & -- & -- \\
L200\_echelle & 1 & 17762 & -- & -- \\
LHIRES \#29 & 1 & 14595 & -- & -- \\
LHIRES 3 & 1 & 7000 & -- & -- \\
LHIRES C11 35um slit 2400l/mm & 4 & 13990, 14220, 14708, 14402 & -- & -- \\
LHIRES III & 17 & 17000, 11600, 17259, 16020, 14577 & -- & -- \\
LHIRES III \#215 & 2 & 17000 & -- & -- \\
LHIRES-A12t & 3 & 6000 & -- & -- \\
LHIRES-B & 86 & 7000, 5000, 4974, 6304, 6020 & -- & -- \\
LHIRES1 & 4 & 6000 & -- & -- \\
LHIRES3 & 47 & 15000, 17000, 11157, 10535, 12305 & -- & -- \\
LHIRES3\#129 & 1 & 13326 & -- & -- \\
LHIRES3\#151 & 1 & 17000 & -- & -- \\
LHIRES3\#194 fente 35 & 4 & 15000 & -- & -- \\
LHIRES3\_2400 & 1 & 16300 & -- & -- \\
LHIRESIII & 7 & 17000, 13399, 13500, 12714, 18704 & -- & -- \\
LHiResIII & 6 & 13500 & -- & -- \\
LIHRES III & 35 & 20000 & -- & -- \\
LIHRES III -1800 & 16 & 11000 & -- & -- \\
LINX (echelle) & 29 & 9000 & -- & -- \\
Lhires III & 1 & 17000 & -- & -- \\
MUSSOL (echelle) & 6 & 9000 & -- & -- \\
NOU\_T (ECHELLE) & 10 & 9000 & -- & -- \\
StarEX & 2 & 18276, 18207 & -- & -- \\
StarEx & 4 & 15951, 14027, 14998, 15955 & -- & -- \\
eShel & 6 & 10000, 11000 & -- & -- \\
eShel101 & 23 & 10000, 11000 & -- & -- \\
eShel\_T\_Garrel & 16 & 11000, 10000 & -- & -- \\
\end{longtable}

\FloatBarrier  

\section{Corner plots and adjustment of $ \rm H\alpha$ line profiles and SED} 

\begin{figure*}
\begin{center}
\includegraphics[scale=0.6]{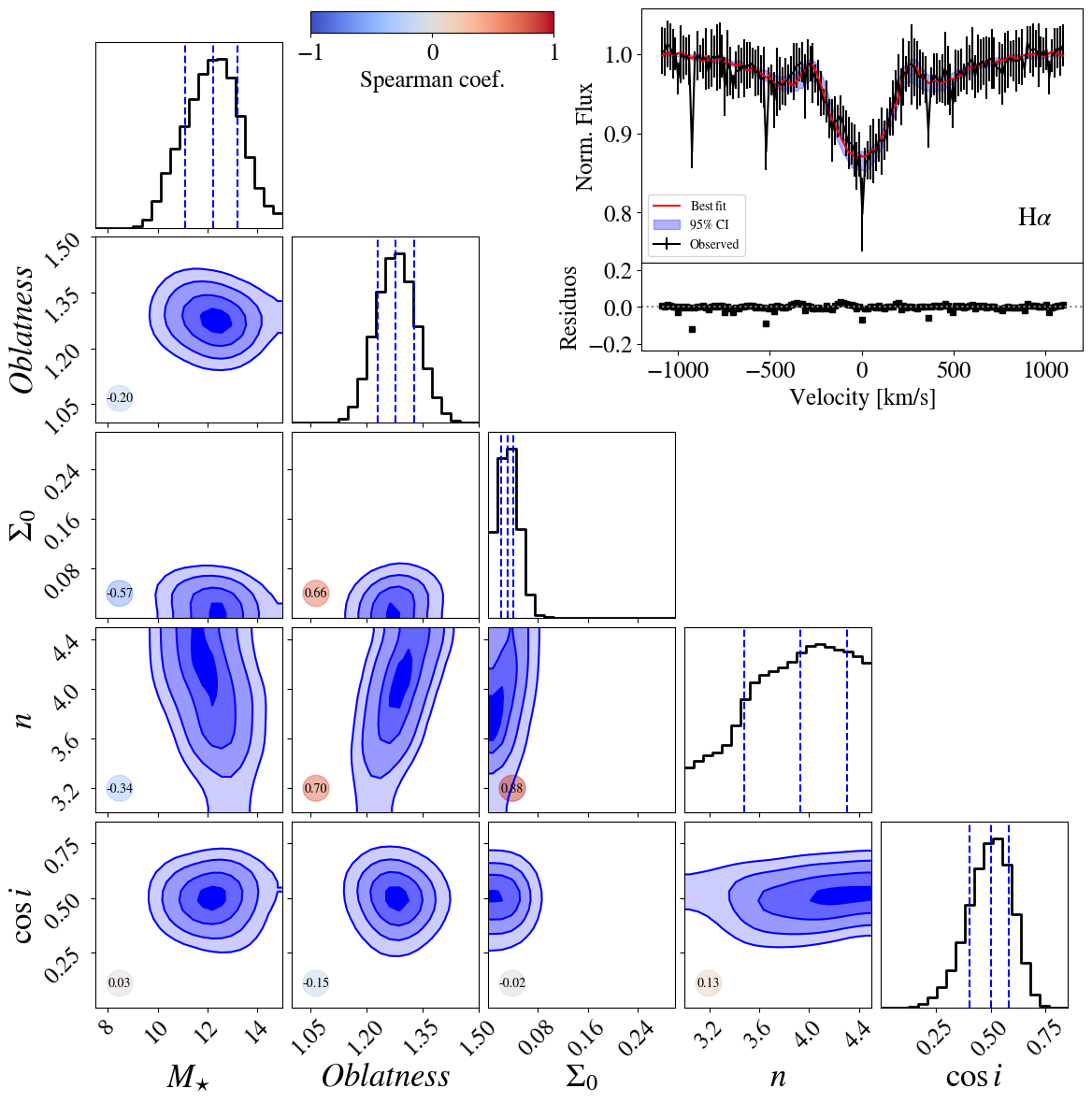}
\caption{Corner plot of $ \rm H\alpha$ line $ \rm \pi$ Aquarii observed in 2001. The 2D histograms for each pair of parameters are shown in blue color. In the main diagonal, the PDFs for each parameter are displayed. The colored circles indicate the Spearman coefficient for the correlation between the pairs of parameters. On the upper right inset, the observational data and residuals are plotted. The red line corresponds to the model that maximizes the likelihood, and the blue region represents the reliability interval of 95\% constructed from 100 random models sampled by the code.}
\label{fig:corner2001}
\end{center}
\end{figure*}

\begin{figure*}
\begin{center}
\includegraphics[scale=0.6]{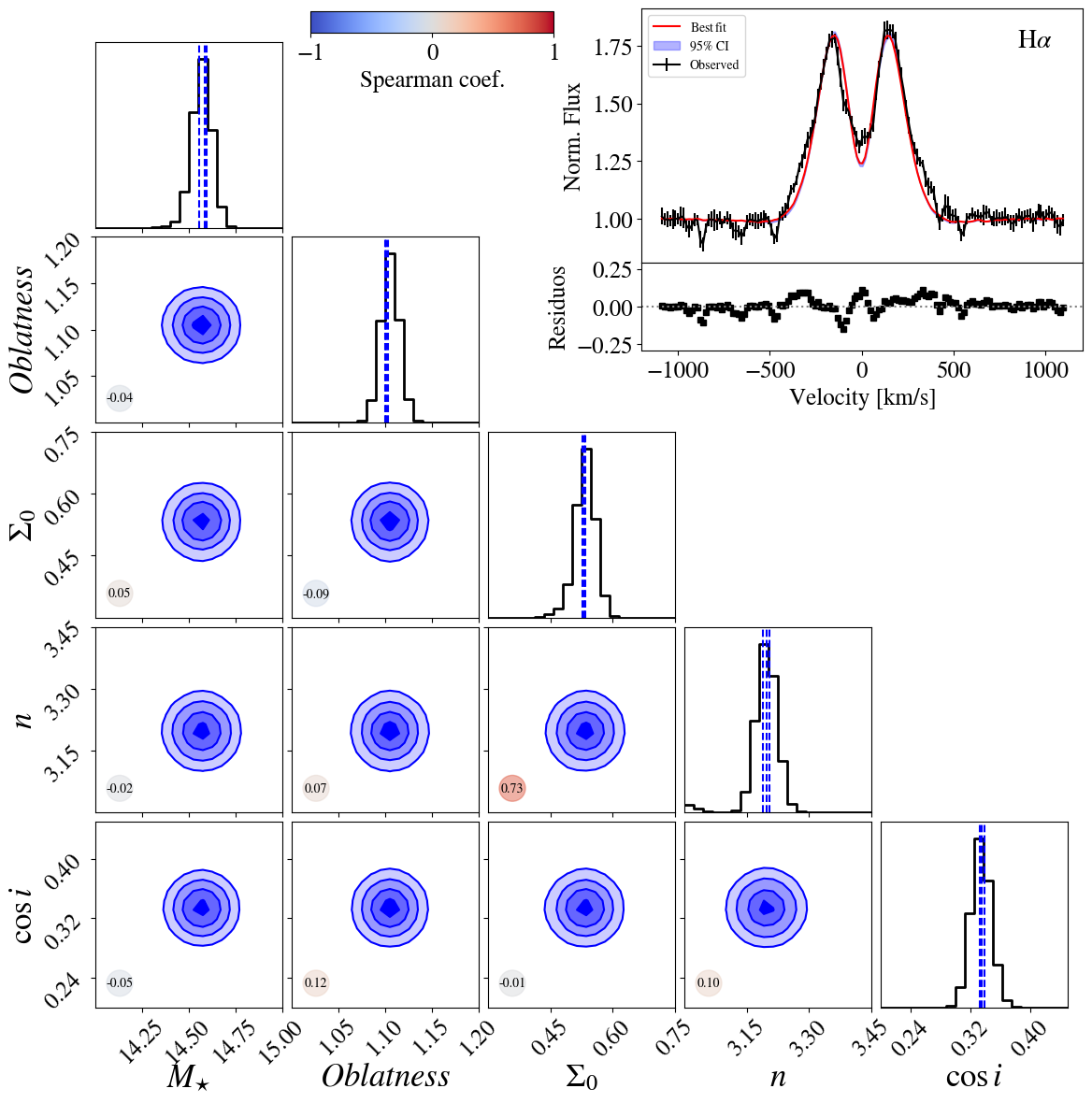}
\caption{Similar to Fig.~\ref{fig:corner2001} but for $ \rm H\alpha$ observed in 2011.}
\label{fig:corner2011}
\end{center}
\end{figure*}

\begin{figure*}
\begin{center}
\includegraphics[scale=0.6]{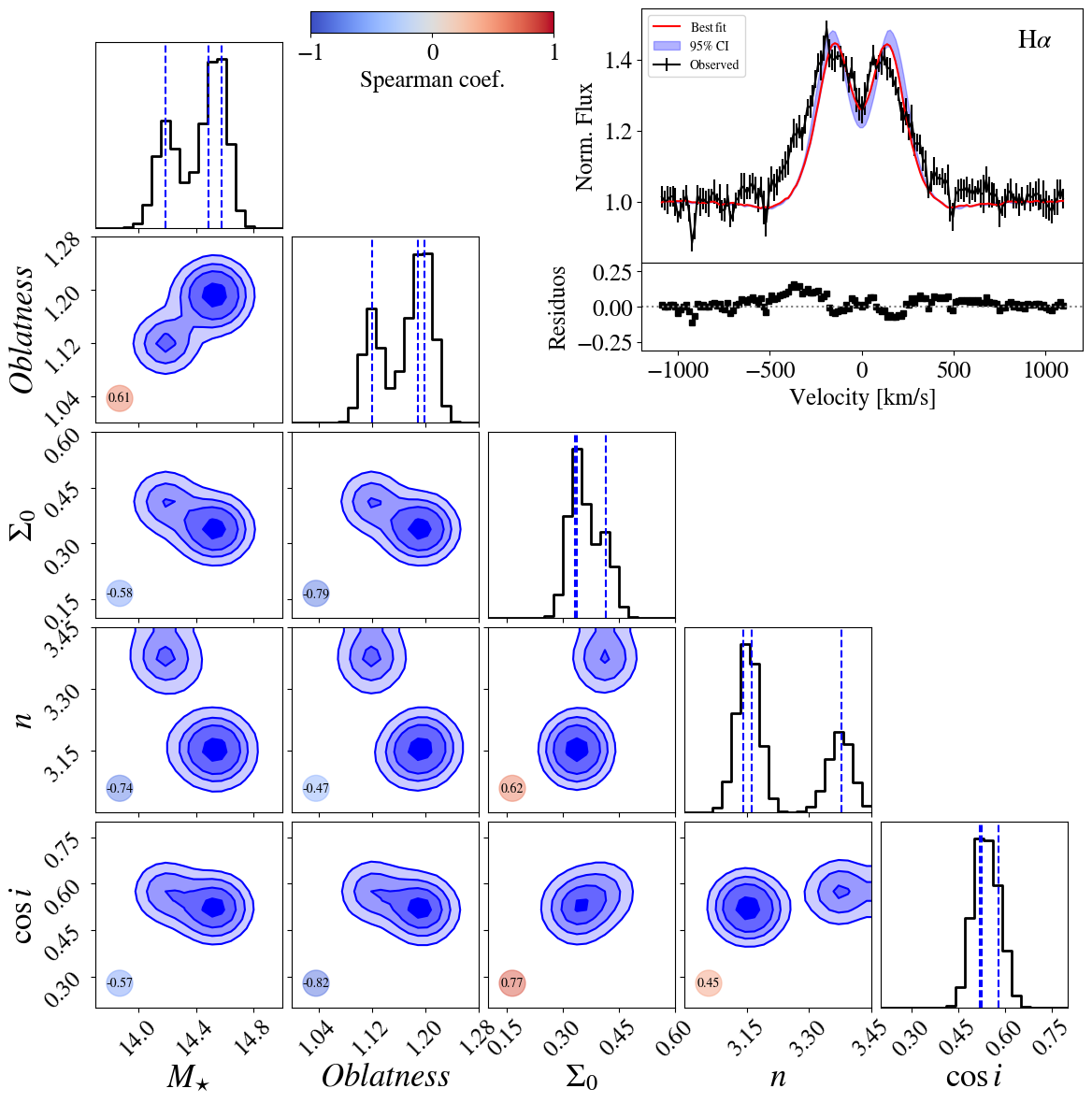}
\caption{Similar to Fig.~\ref{fig:corner2001} but for $ \rm H\alpha$ observed in 2014.}
\label{fig:corner2014}
\end{center}
\end{figure*}

\begin{figure*}
\begin{center}
\includegraphics[scale=0.6]{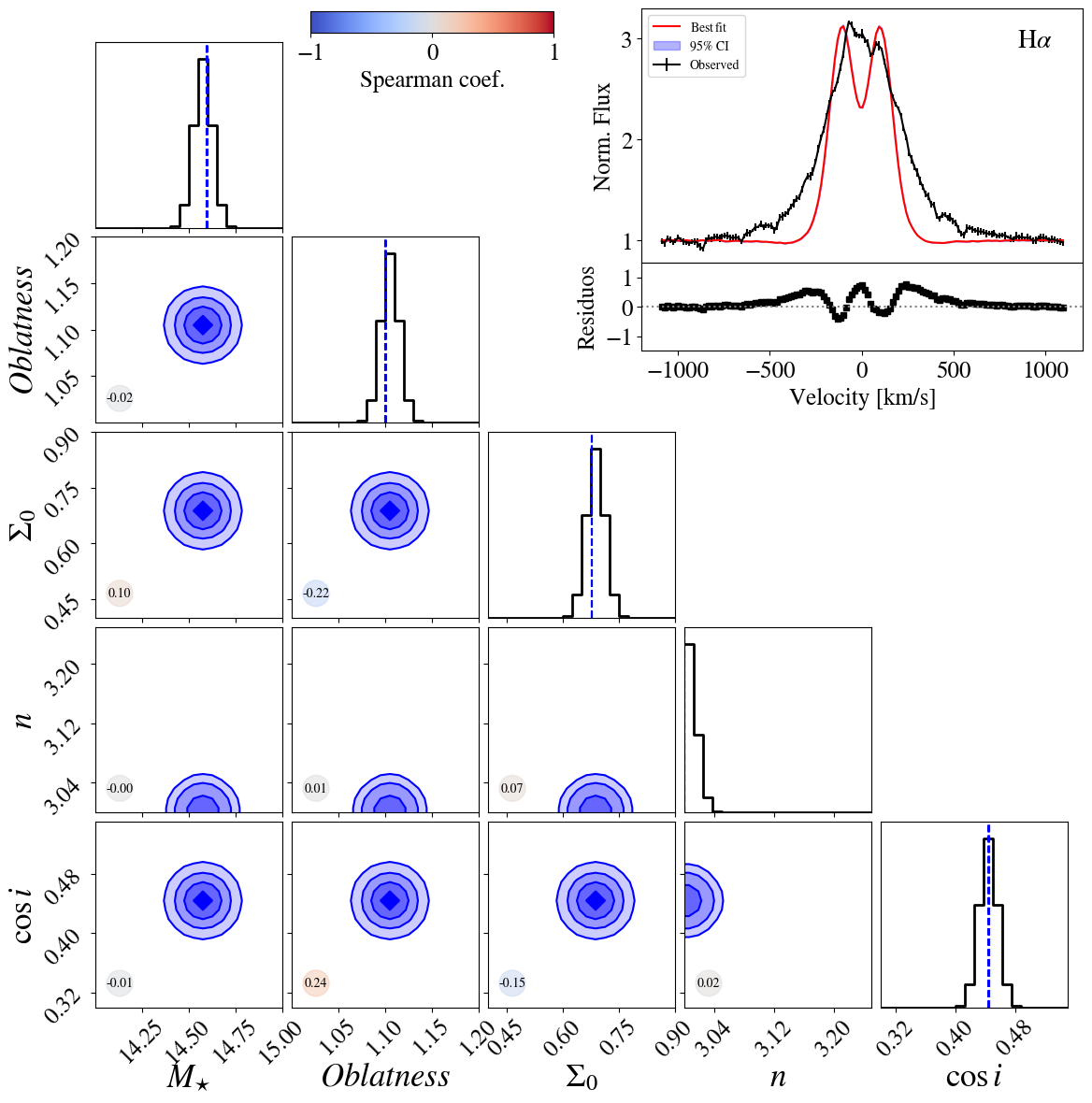}
\caption{Similar to Fig.~\ref{fig:corner2001} but for $ \rm H\alpha$ observed in 2018.}
\label{fig:corner2018}
\end{center}
\end{figure*}

\begin{figure*}
\begin{center}
\includegraphics[scale=0.6]{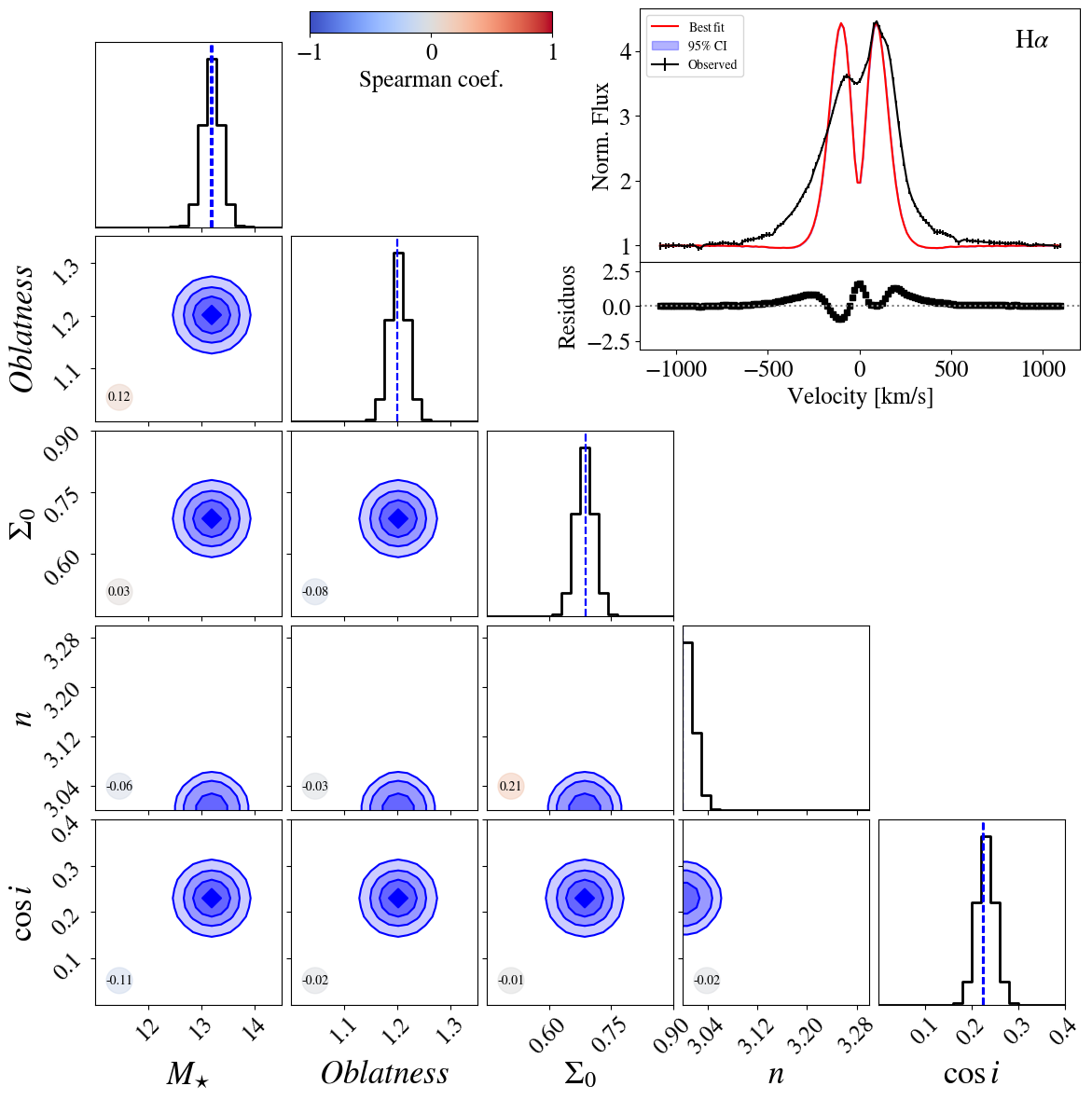}
\caption{Similar to Fig.~\ref{fig:corner2001} but for $ \rm H\alpha$ observed in 2022.}
\label{fig:corner2022}
\end{center}
\end{figure*}

\begin{figure*}
\begin{center}
\includegraphics[scale=0.4]{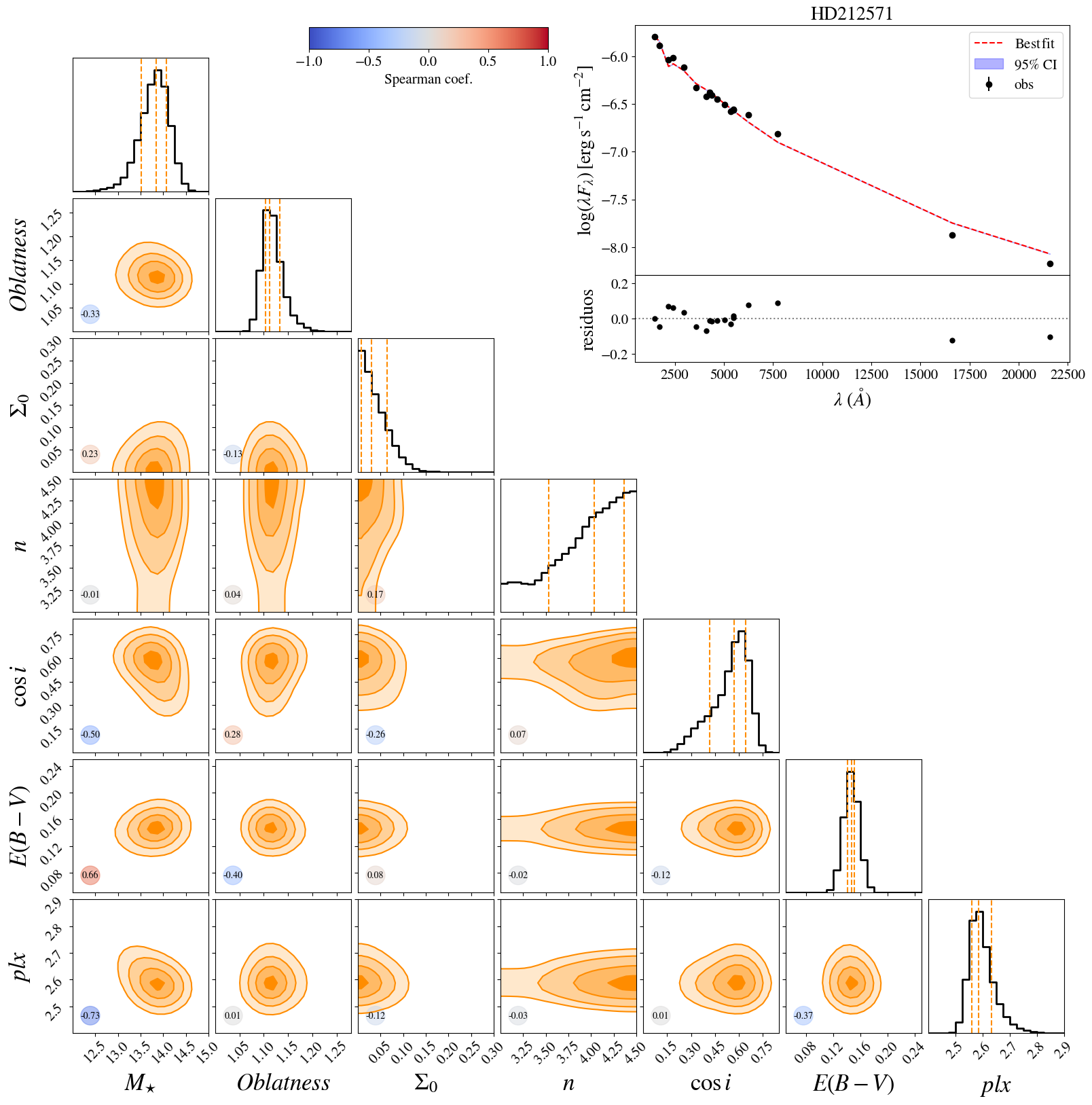}
\caption{Corner plot of SED $ \rm \pi Aquarii $. The $ \rm E(B-V)$ and parallax were also modeled. The 2D histograms for each pair of parameters are shown in dark orange. In the main diagonal, the PDFs for each parameter are displayed. The colored circles indicate the Spearman coefficient for the correlation between the pairs of parameters. In the upper-right inset, the observational data are in black dots, and residuals are plotted. The dashed red line corresponds to the model that maximizes the likelihood, and the blue region represents the reliability interval of 95\% constructed from 100 random models sampled by the code.}
\label{fig:cornersed}
\end{center}
\end{figure*}

\FloatBarrier  
    
\section{Evolution of lines in time}

\begin{figure*}[ht!]
\begin{center}
\includegraphics[scale=0.48]{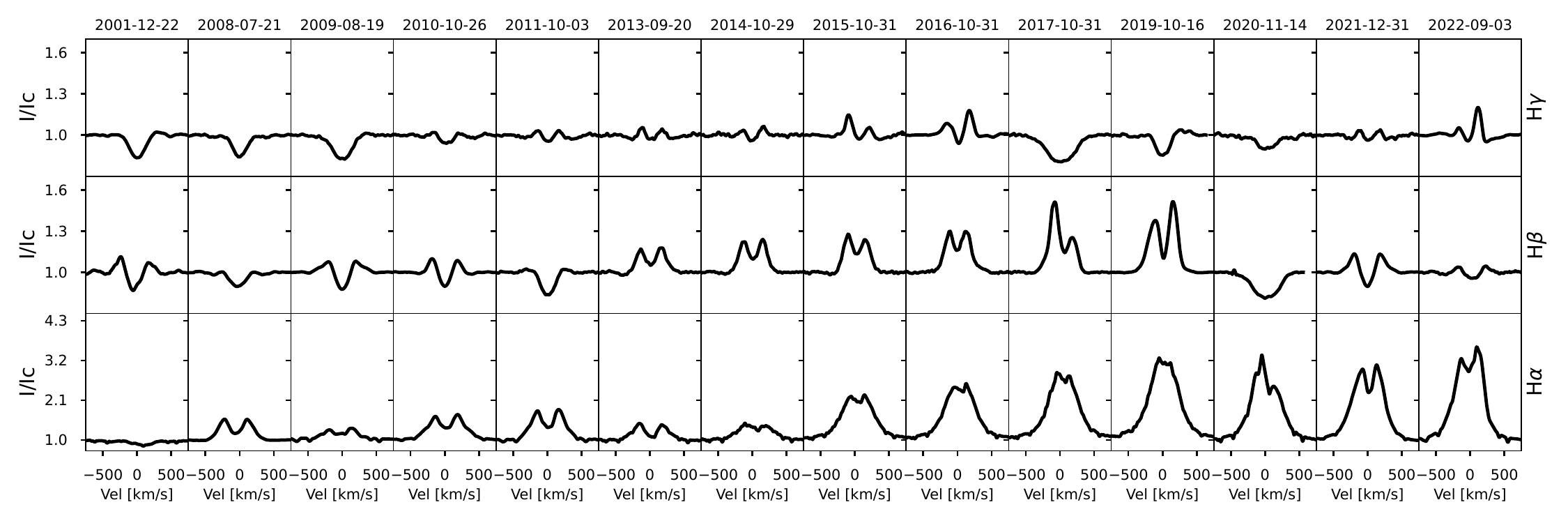}
\caption{Comparison of the evolution of Balmer lines H$\alpha$, H$\beta$, and H$\gamma$ for the same observation dates.}
\label{app:evolbalmer}
\end{center}
\end{figure*}

\begin{figure*}[ht!]
\begin{center}
\includegraphics[scale=0.48]{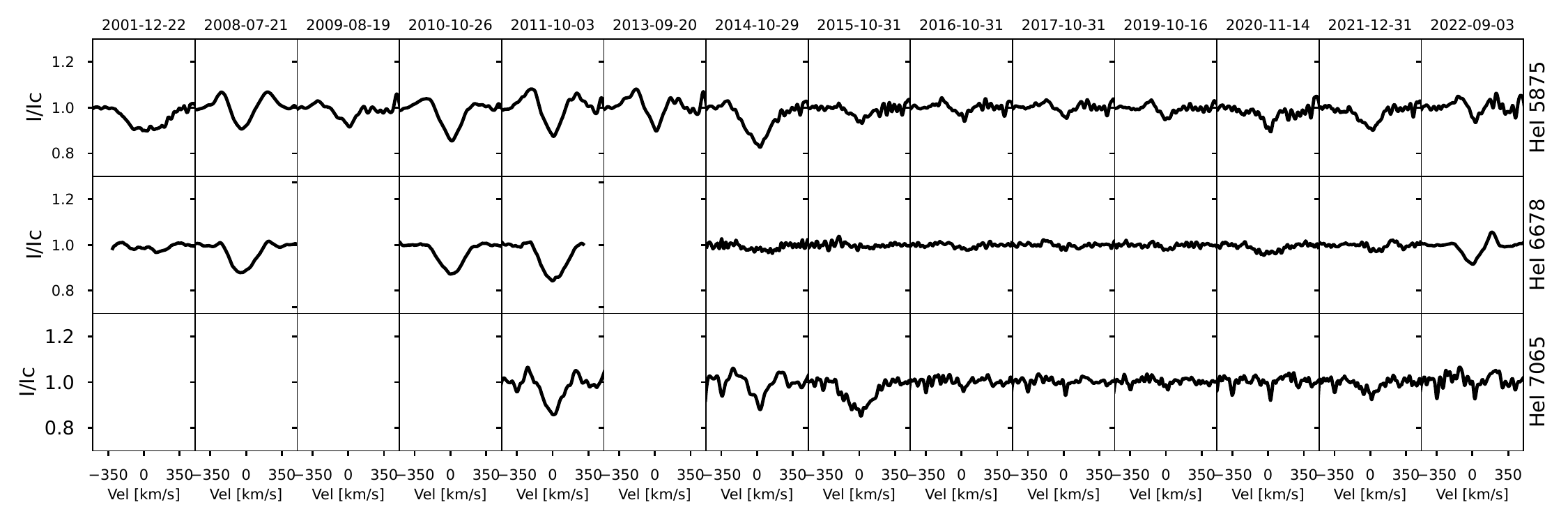}
\caption{Same as Fig.~\ref{app:evolbalmer} but for helium lines.}
\label{app:evolhelium}
\end{center}
\end{figure*}

\begin{figure*}[ht!]
\begin{center}
\includegraphics[scale=0.48]{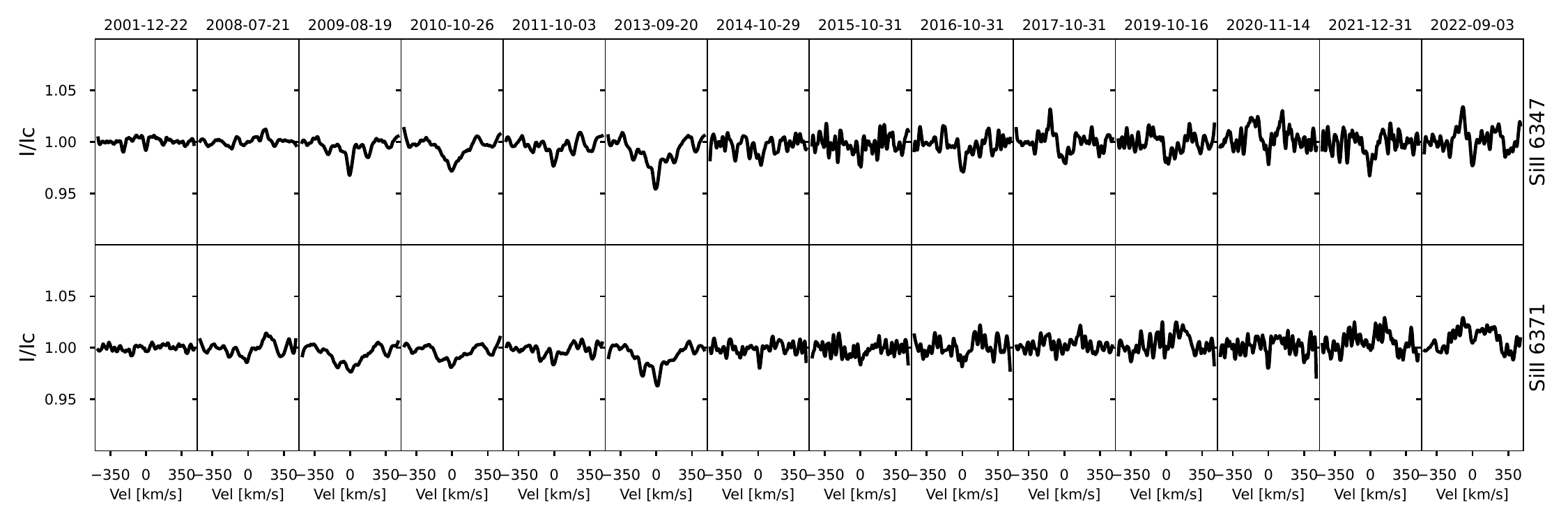}
\caption{Same as Fig.~\ref{app:evolbalmer} but for silicon lines.}
\label{app:evolsi}
\end{center}
\end{figure*}

\begin{figure*}[ht!]
\begin{center}
\includegraphics[scale=0.5]{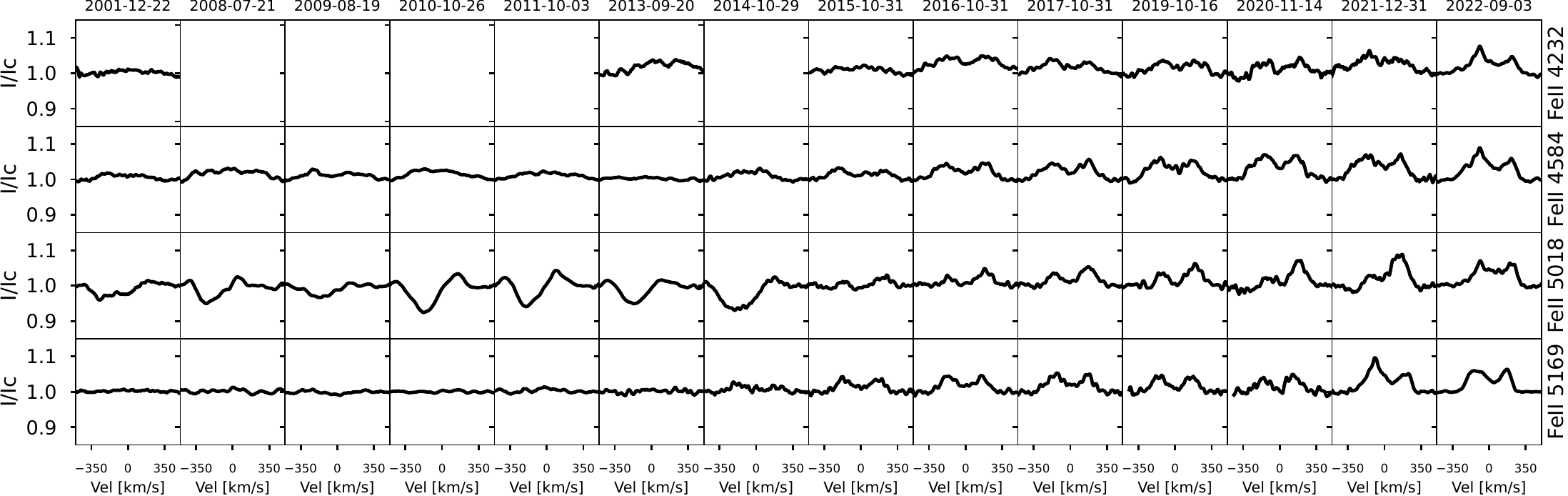}
\includegraphics[scale=0.5]{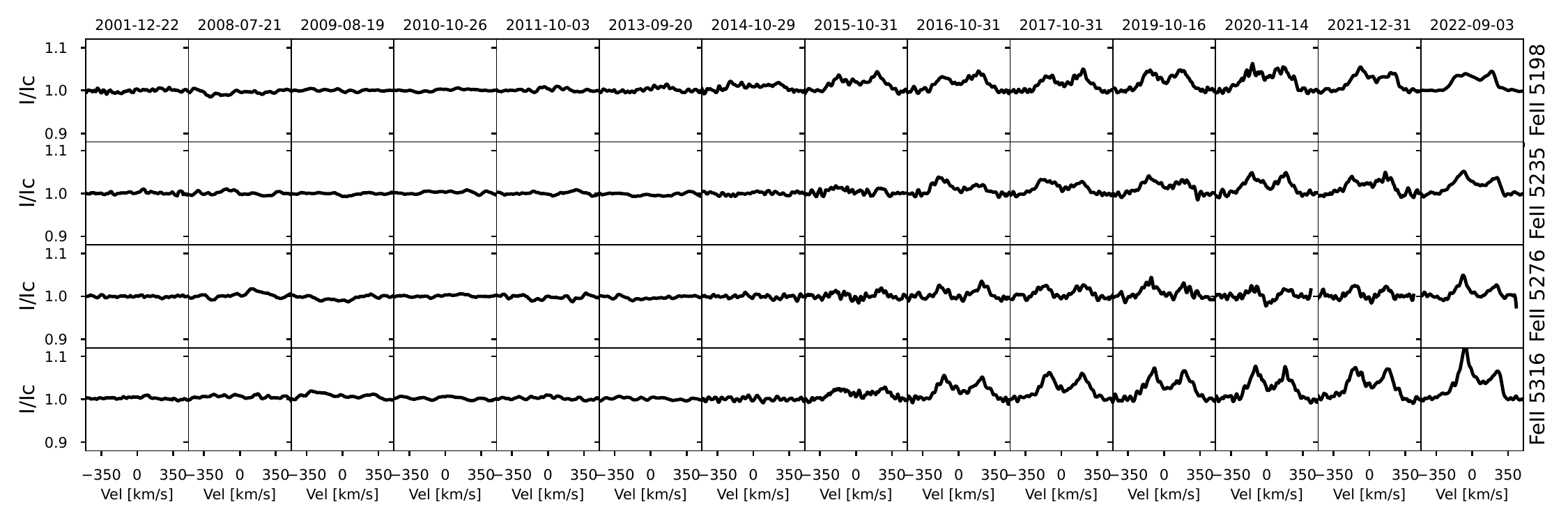}
\includegraphics[scale=0.5]{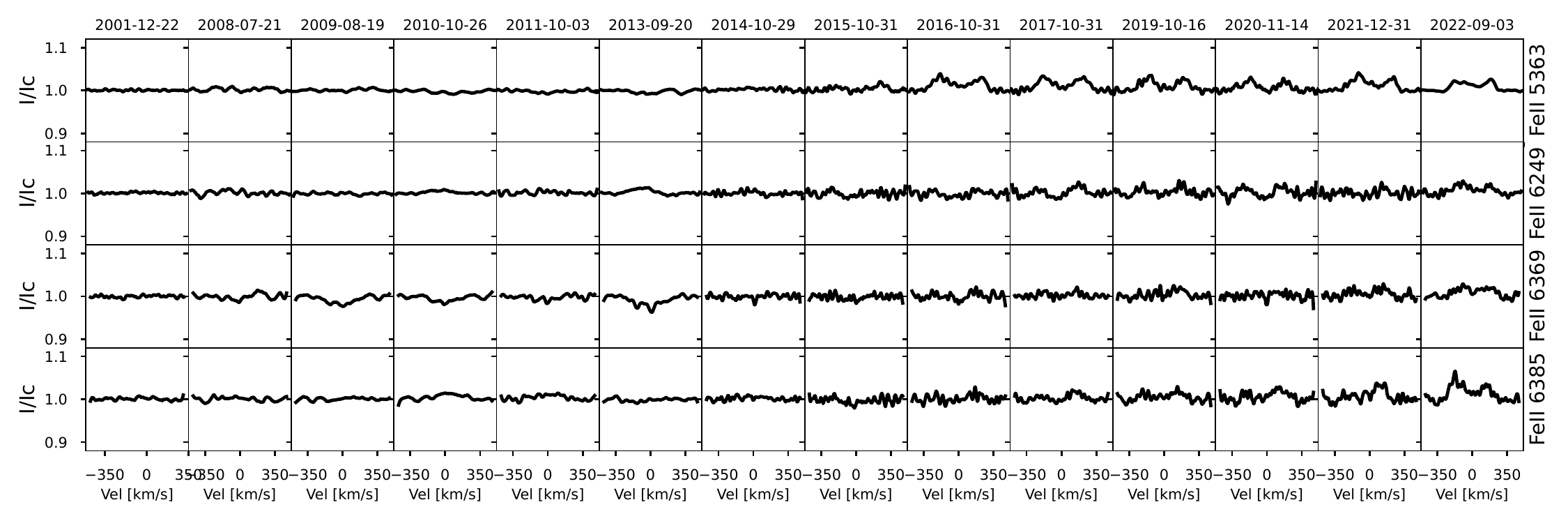}
\caption{Same as Fig.~\ref{app:evolbalmer} but for iron lines.}
\label{app:evoliron1}
\end{center}
\end{figure*}

\end{appendix}
\end{document}